\documentclass[12pt,a4paper]{article}
\usepackage[english]{babel}
\usepackage[T1]{fontenc}
\usepackage{ae,aecompl}
\usepackage[numbers,comma,square,compress]{natbib}
\usepackage{amsmath,amssymb,amsfonts}
\usepackage{bm}
\usepackage{ifpdf}
\ifpdf
  \usepackage[pdftex,a4paper,hmargin=25mm,vmargin=25mm]{geometry}
  \usepackage[pdftex,bookmarks=true]{hyperref}
\else
  \usepackage[dvips,a4paper,hmargin=25mm,vmargin=25mm]{geometry}
  \usepackage[dvips,bookmarks=true]{hyperref}
\fi

\numberwithin{equation}{section}

\newcommand{\email}[1]{\href{mailto:#1}{#1}}
\newcommand{\nn}{\nonumber}

\newcommand{\p}{\partial}
\newcommand{\pb}[1]{\left\{#1\right\}}
\newcommand{\cM}{\mathcal{M}}
\newcommand{\cH}{\mathcal{H}}
\newcommand{\cL}{\mathcal{L}}
\newcommand{\cF}{\mathcal{F}}

\newcommand{\cP}{\mathcal{P}}
\newcommand{\cA}{\mathcal{A}}
\newcommand{\bx}{\bm{x}}
\newcommand{\by}{\bm{y}}
\newcommand{\bz}{\bm{z}}
\DeclareMathOperator{\sgn}{sgn}
\newcommand{\gM}[1][4]{{}^{(#1)}\!g}
\newcommand{\RM}[1][4]{{}^{(#1)}\!R}
\newcommand{\GM}[1][4]{{}^{(#1)}\!G}
\newcommand{\GammaM}[1][4]{{}^{(#1)}\Gamma}
\newcommand{\projector}[2]{g^{#1}_{\phantom{#1}#2}}
\newcommand{\Proj}[2]{P_{#1}^{\phantom{#1}#2}}

\begin{document}

\begin{center}
{\Large Higher derivative gravity with spontaneous\\
symmetry breaking: Hamiltonian analysis\\
\vspace{.3em}
 of new covariant renormalizable gravity}\\
\vspace{1em}
Masud Chaichian$\,^1$,
Josef Kluso\v{n}$\,^2$,
Markku Oksanen$\,^1$,
Anca Tureanu$\,^1$\,\footnote{Email addresses:
\email{masud.chaichian@helsinki.fi} (M. Chaichian),
\email{klu@physics.muni.cz} (J. Kluso\v{n}),
\email{markku.oksanen@helsinki.fi} (M. Oksanen),
\email{anca.tureanu@helsinki.fi} (A. Tureanu)}\\
\vspace{1em}
$^1$\textit{Department of Physics, University of Helsinki, P.O. Box
64,\\ FI-00014 Helsinki, Finland}\\
$^2$\textit{Department of Theoretical Physics and Astrophysics, Faculty
of Science,\\
Masaryk University, Kotl\'a\v{r}sk\'a 2, 611 37, Brno, Czech Republic}
\end{center}

\begin{abstract}
In order to explore some general features of modified theories of
gravity which involve higher derivatives and spontaneous Lorentz and/or
diffeomorphism symmetry breaking, we study the recently proposed new
version of covariant renormalizable gravity (CRG). CRG attains
power-counting renormalizability via higher derivatives and
introduction of a constrained scalar field and spontaneous symmetry
breaking. We obtain an Arnowitt-Deser-Misner representation of the
CRG action in four-dimensional spacetime with respect to a foliation of
spacetime adapted to the constrained scalar field. The resulting action 
is analyzed by using Hamiltonian formalism. We discover that CRG
contains two extra degrees of freedom. One of them carries negative
energy (a ghost) and it will destabilize the theory due to its
interactions. This result is in contrast with the original paper [Phys.
Lett. B \textbf{701}, 117 (2011), arXiv:1104.4286 [hep-th]], where it
was concluded that the theory is free of ghosts and renormalizable when
we analyze fluctuations on the flat background.
\end{abstract}

\vspace{.5em}
\begin{tabular}{ll}
\textbf{PACS}:& 04.50.Kd (Modified theories of gravity),
04.60.-m (Quantum gravity),\\
& 11.10.Ef (Lagrangian and Hamiltonian approach),\\
& 98.80.Cq (Particle-theory and field-theory models of the early
Universe)
\end{tabular}

\section{Introduction}

In the recent years modified theories of gravity have attracted a
considerable amount of attention. These modifications of general
relativity (GR) aim to improve the behavior of the theory either at
high energies or at large distances. This is motivated by the fact that
although GR is very successful in describing gravitational phenomena in
intermediate distances, it has become increasingly evident that it may
need to be completed both at the high energy realm (quantum gravity and
renormalizability) and at large distances (dark energy and dark matter)
in order to achieve a more plausible theoretical framework and also
more convincing agreement with observational data. The fact that
according to standard cosmology only four percent of the energy content
of the universe has been observed by means other than gravity is a huge
gap in our understanding on the nature of gravity.

Perturbative renormalization of GR requires us to include invariants
quadratic in curvature into Lagrangian as counterterms
\cite{tHooft:1974,Deser:1974}. The Riemann tensor squared term can be
excluded in 4-dimensional spacetime due to the Gauss-Bonnet topological
invariance. A gravitational Lagrangian consisting of the scalar
curvature $R$, scalar curvature squared $R^2$, and Ricci tensor squared
$R_{\mu\nu}R^{\mu\nu}$ terms is indeed renormalizable via dimensional
regularization \cite{Stelle:1977}. Unfortunately, it includes a massive
spin-2 excitation with negative energy \cite{Stelle:1977,Stelle:1978},
because $R_{\mu\nu}$ includes second-order time derivatives of every
component of the metric. This means the theory suffers from
Ostrogradskian instability \cite{Woodard:2007}, namely the massive
spin-2 ghost destabilizes the theory, an illness that hampers many
higher derivative theories. Therefore much interest has been
directed to modified gravity with the Lagrangian $R+\alpha R^2$
\cite{Starobinsky:1980,Strominger:1984},
which contains only one extra scalar degree of freedom and it is healthy
for $\alpha>0$. However, the $R^2$ term alone is not sufficient for
renormalizability. More generally one considers $f(R)$ gravity whose
Lagrangian is a nonlinear function of the scalar curvature. For a recent
review and references, see for example \cite{Sotiriou:2010}. These
theories have the advantage of being able to realize cosmological
phases of accelerated expansion (inflation and present era) without
additional dark components. Indeed $f(R)$ gravity has an equivalent
representation as a minimally coupled scalar-tensor theory and therefore
it is practically equivalent to quintessence in this respect. Other
well-known examples of modified theories of gravity are for example
Brans-Dicke theory \cite{Brans:1961}, another scalar-tensor theory, and
the relativistic tensor-vector-scalar implementation
\cite{Bekenstein:2004} of modified Newtonian dynamics
\cite{Milgrom:1983}. It appears that such modifications do not
eventually make the theory renormalizable.

Ho\v{r}ava-Lifshitz (HL) gravity \cite{Horava:2009uw} is a
power-counting renormalizable field theory of gravity which is based on
the idea that space and time scale differently at high energies,
\begin{equation}
\bx \rightarrow b \bx \,,\qquad t \rightarrow b^z t \,,
\end{equation}
with a dynamic critical exponent $z$. This enables one to modify the
ultraviolet behavior of the graviton propagator to $|\bm{k}|^{-2z}$,
where $\bm{k}$ is the spatial momentum. In $D$ spatial dimensions,
choosing $z=D$ in the ultraviolet fixed point ensures the gravitational
constant is dimensionless and the theory is power-counting
renormalizable. Such a spacetime admits a preferred foliation into
spatial hypersurfaces, and hence the local Lorentz invariance is
broken. General diffeomorphism invariance of GR is broken down to 
foliation-preserving diffeomorphisms, given in the infinitesimal
form as
\begin{equation}\label{fp-diffeomorphism}
\delta t = f(t) \,,\qquad \delta\bx = \bm{\xi}(t,\bx) \,.
\end{equation}
The reduced symmetry means HL gravity contains an extra scalar degree of
freedom. In the original versions of HL gravity, and also in the
extensions without a so-called detailed balance condition
\cite{Sotiriou:2009}, the extra scalar degree of freedom does not
decouple in the infrared limit. Instead it becomes strongly coupled
\cite{Charmousis:2009,Blas:2009,Koyama:2010}, which means GR cannot
be recovered in the low energy limit and also that renormalizability
could be ruined. In order to cure this problem, an extension of HL
gravity was proposed in \cite{Blas:2010a} (also see
\cite{Blas:2010b,Kimpton:2010}), where the potential part of the action
is extended with terms that contain spatial derivatives of the lapse
function $N$ as the vector
\begin{equation}\label{a_i}
a_i=\frac{\p_iN}{N}\,.
\end{equation}
As a result the extra scalar mode attains a healthy quadratic action
and no strong coupling appears, assuming certain parameters are chosen
to be sufficiently small. Even this ``healthy version'' of HL gravity
might still run into trouble with tests of the equivalence principle
\cite{Padilla:2010}. Another possible way to deal with the extra scalar
mode is to extend the symmetry group of HL gravity so that the
extra scalar degree of freedom is eliminated. This approach was taken in
 \cite{Horava:2010zj}, where an extra local $U(1)$ symmetry is
introduced, which may eliminate the extra scalar mode. Whether this
truly cures the problems of original HL gravity is not quite clear yet;
for possible problems and ways out see
\cite{daSilva:2010,Sotiriou:2011}.

More general theories of the HL type have been proposed in
\cite{Chaichian:2010a,Carloni:2010}, also exploring the challenge of
long-distance behavior of gravity.

Recently, the so-called covariant renormalizable gravity (CRG) was
proposed \cite{Nojiri:2010b}. CRG aims to provide a power-counting
renormalizable field theory of gravity that is covariant under spacetime
diffeomorphism and possesses local Lorentz invariance at the fundamental
level. CRG aims to achieve a similar ultraviolet behavior of the
graviton propagator as HL gravity, but without introducing explicitly
Lorentz noninvariant terms into the action. Lorentz invariance of the
graviton propagator of CRG is, however, broken spontaneously at high
energies. This is achieved by introducing a scalar field, which is
coupled to spacetime in a rather complicated way, and a constraint on
the scalar field that breaks Lorentz symmetry spontaneously.
Since the CRG action contains higher-order derivatives it exhibits extra
degrees of freedom. A new version of CRG has been proposed
\cite{Kluson:2011rs}, where a perturbative analysis around Minkowski
spacetime showed that the extra degrees of freedom present in the
theory do not propagate.
However, we should note that the renormalizability of CRG, as well as of
the HL theory, is assumed only based on the power-counting arguments.
There are several potential pathologies that could ruin the
renormalizability of this theory, such as gradient instabilities,
ghosts, or strong coupling.
Since a violation of Lorentz invariance has never been observed, one
could try to argue that CRG is a more natural modification of GR than
the explicitly Lorentz noninvariant ones, in particular HL gravity and
its generalizations. On the other hand, Lorentz invariance could equally
well be broken explicitly at very high energies as long as it is somehow
restored at sufficiently low energies.

At present, most modified theories of gravity should be treated as
effective or phenomenological theories, since they are not derived from
any compelling first principles, rather constructed to meet some
specific purposes. Nevertheless these theories can teach us a great deal
about the aspects of gravity, and ultimately help us in laying down the
foundations for the next paradigm of space, time and gravity.

Hamiltonian formalism provides a powerful tool for the analysis of
constrained systems such as gravity. The number and nature of
constraints and physical degrees of freedom is not always evident from
the Lagrangian. Even crucial inconsistencies can sometimes be found.
For example, it has been shown by Hamiltonian analysis that the
original version of HL gravity is physically inconsistent at high
energies \cite{Henneaux:2010}, unless the projectability condition is
imposed on the lapse function $N$ ($N$ must depend on time only).
A similar result regarding the projectability of $N$ has also been
obtained for the more general modified $F(R)$ HL gravity
\cite{Chaichian:2010b}.
When the potential part of the action is extended with terms involving
the vector \eqref{a_i} \cite{Blas:2010a} in order to cure the strong
coupling problem, the theory also gains a consistent Hamiltonian
structure \cite{Kluson:2010nf,Donnelly:2011}.

In a previous paper \cite{Chaichian:2011sx} we studied the first
version of CRG from the point of view of Hamiltonian analysis. It was
found that there indeed exist extra degrees of freedom compared to GR,
because there are not enough constraints to eliminate them. In this
paper we analyze the new version of CRG \cite{Kluson:2011rs},
concentrating on the renormalizable model in 4-dimensional spacetime.
In particular we are interested in whether the new model contains
extra modes similarly as original CRG and whether the possible extra
modes are pathological ghosts. We shall see that the results of this
study will be of interest to other conceivable generally covariant
higher derivative theories of gravity which aim to achieve
power-counting renormalizability via spontaneous (constraint induced)
Lorentz and/or diffeomorphism symmetry breaking.

First the action of new CRG is introduced in Sec.~\ref{sec2}.
We obtain the Arnowitt-Deser-Misner (ADM) representation of the new CRG
action with respect to a foliation of spacetime adapted to the
constrained scalar field in Sec.~\ref{sec3}. The resulting action
contains time derivative of the extrinsic curvature and powers of the
extrinsic curvature up to sixth order. In Sec.~\ref{sec4}, introducing
some additional fields enables us to obtain a first-order action with a
kinetic part quadratic in extrinsic curvature and other first time
derivatives. Certain solutions of the first-order action are discussed
in Sec.~\ref{sec5}. In Sec.~\ref{sec6}, we analyze the action using
Hamiltonian formalism. A few alternative sets of variables for
Hamiltonian formulation are considered in Sec.~\ref{sec7}. Conclusions
and some further discussion are presented in Sec.~\ref{sec8}.

\section{Action}\label{sec2}

We consider the new version of covariant renormalizable gravity with
projectors \cite{Kluson:2011rs}. For definiteness we shall consider
the specific model corresponding to critical exponent $z=3$ which should
be power-counting renormalizable in 4-dimensional spacetime.
The action reads
\begin{multline}\label{S3}
S_3 = \int d^4 x \sqrt{-\gM} \left[ \frac{\RM}{2\kappa^2} - \alpha
\Proj{\alpha}{\mu}\Proj{\beta}{\nu} \left( \RM_{\mu\nu} -
\frac{1}{2U_0}\p_\rho \phi \nabla^\rho \nabla_\mu \nabla_\nu \phi
\right) \right.\\
\times \left( \p^\mu \phi \p^\nu \phi \nabla_\mu \nabla_\nu - \p_\mu
\phi \p^\mu \phi \nabla^\nu \nabla_\nu \right) P^{\alpha\mu}P^{\beta\nu}
\left( \RM_{\mu\nu} - \frac{1}{2U_0}\p_\rho \phi \nabla^\rho \nabla_\mu
\nabla_\nu \phi \right) \\
- \left. \lambda \left( \frac{1}{2}\p_\mu \phi \p^\mu \phi + U_0 \right)
\right] \,,
\end{multline}
where the projector is defined by
\begin{equation}\label{scalarprojector}
\Proj{\alpha}{\mu} = \delta_\alpha^\mu + \frac{\p_\alpha \phi \p^\mu
\phi}{2U_0} \,.
\end{equation}
Variation of the action \eqref{S3} with respect to the Lagrange
multiplier $\lambda$ implies the constraint on the scalar field $\phi$
as
\begin{equation}\label{phiconstraint}
\frac{1}{2}\p_\mu \phi \p^\mu \phi + U_0 = 0 \,.
\end{equation}
For simplicity we consider $U_0$ to be a positive constant, as was
considered in the original proposal \cite{Kluson:2011rs}, but more
generally it could be any positive function of $\phi$. The constraint
\eqref{phiconstraint} means the gradient $\p^\mu \phi$ of the scalar
$\phi$ is timelike everywhere.

For long distances the behavior of CRG is supposed to be dominated by
the Einstein-Hilbert part of the action \eqref{S3}, producing physics
consistent with GR. For short distances and high energies the action of
CRG is dominated by the higher-derivative terms with the coupling
$\alpha$. These higher-derivative terms enable the power-counting
renormalizability. Assuming there are no pathologies in the theory,
Solar System and laboratory tests could be used to bound the coupling
$\alpha$ to sufficiently small values. However, the action \eqref{S3} is
of uncommon type and it is not at all clear what kind of physical
degrees of freedom it contains, because of the higher-order time
derivatives, the constraint on the scalar field, and the complicated
couplings between the fields. The action indeed contains higher-order
time derivatives of both the metric and the scalar field, which could
cause serious problems. In particular, the existence of unstable ghosts
is possible, because every extra time derivative of a variable
translates into an extra physical degree of freedom, provided the number
of constraints in the system is unaltered, and such a higher-order
degree of freedom carries an energy with opposite sign compared to the
corresponding lower-order degree of freedom. Therefore a close
inspection of the action is required in order to see whether it defines
a healthy theory in the first place.

Let us see how the argument for power-counting renormalizability arises.
Choosing a solution of the equation of motion \eqref{phiconstraint}
spontaneously breaks the Lorentz invariance and/or the full general
covariance of the action \eqref{S3} \cite{Kluson:2011rs}. According
to Eq.~\eqref{phiconstraint} $\p^\mu\phi$ is timelike and
hence one can choose the direction of time to be parallel to
$\p_\mu\phi$ at least locally. Then one finds that the higher-derivative
modification added to Einstein-Hilbert action in \eqref{S3}, i.e., the
term with the coupling constant $\alpha$, turns out to contain only
spatial derivatives of the perturbation $h_{\mu\nu}$ to Minkowski
metric $\eta_{\mu\nu}$; the nearly flat metric of spacetime is given as
$\gM_{\mu\nu}=\eta_{\mu\nu}+h_{\mu\nu}$.
In the present case ($z=3$) there are six spatial derivatives of
$h_{\mu\nu}$, while in the general case there are $2z$ spatial
derivatives, yielding the desired $|\bm{k}|^{-2z}$ modification of the
graviton propagator. Power-counting renormalizability is achieved when
$z$ is equal to the dimensionality of space.
The form of this modification to GR was specifically constructed so that
time derivatives of $h_{\mu\nu}$ cancel out in the action once the
Lorentz symmetry is spontaneously broken and the direction of time
has been fixed. This means that all the problems of higher time
derivative theories can be avoided in this linearized and gauge fixed
formulation, quite similarly as in HL gravity where higher time
derivatives are explicitly excluded in the definition of action.
Studying the graviton propagator in \cite{Kluson:2011rs} revealed that
the flat vacuum is unstable for $\alpha>0$, while for $\alpha<0$ no
such problem appeared.

Although we concentrate on the renormalizable case $z=3$ in four
dimensions, actions corresponding to other (higher) values of $z$ could
be considered in a similar fashion by using the results we will obtain
in the following sections.

\section{Arnowitt-Deser-Misner representation}\label{sec3}

We consider the ADM decomposition of the gravitational field
\cite{Arnowitt:1962}; for reviews and mathematical background, see
\cite{ADMreview}. Spacetime is assumed to admit a foliation
into spacelike hypersurfaces $\Sigma_t$ of constant $t$.
The metric of spacetime can be written as
\begin{equation}
\gM_{\mu\nu} = g_{\mu\nu}-n_\mu n_\nu \,,
\end{equation}
where $g_{\mu\nu}$ is the induced metric on the spacelike hypersurfaces
$\Sigma_t$ and $n_\mu$ is the future-directed unit normal to $\Sigma_t$.
The ADM variables consist of the lapse function $N$, the
shift vector $N^i$ and the spatial metric $g_{ij}$ ($i,j=1,2,3)$.
The unit normal to $\Sigma_t$ can be written in terms of the ADM
variables as
\begin{equation}\label{normal}
n_\mu = -N\nabla_\mu t = (-N,\bm{0}) \,,\qquad
n^\mu = \left(n^0, n^i\right)
= \left(\frac{1}{N}, -\frac{N^i}{N}\right) \,.
\end{equation}
Thus the ADM representation for the metric of spacetime reads
\begin{equation}\label{ADMmetric}
\gM_{00} = -N^2+N_i N^i \,,\qquad \gM_{0i} = \gM_{i0} = N_i
\,,\qquad
\gM_{ij} = g_{ij} \,,
\end{equation}
where $N_i=g_{ij}N^j$. Contravariant components of the metric of
spacetime are
\begin{equation}\label{ADMinvmetric}
\gM^{00} = -\frac{1}{N^2} \,,\qquad \gM^{0i}= \gM^{i0} = \frac{N^i}{N^2}
\,,\qquad \gM^{ij} = g^{ij}-\frac{N^i N^j}{N^2} \,,
\end{equation}
where $g^{ij}g_{jk}=\delta^i_k$.
The extrinsic curvature of the spatial hypersurface $\Sigma_t$ is
defined by
\begin{equation}\label{K_ij}
K_{ij} = \frac{1}{2N}\left( \dot{g}_{ij} - 2D_{(i}N_{j)} \right)
\,,\qquad K=g^{ij}K_{ij} \,,
\end{equation}
where the dot denotes the derivative with respect to time $t$.
Quantities defined on the spacetime $\cM$ and associated with its metric
$\gM_{\mu\nu}$ are marked with the prefix ${}^{(4)}$. We denote the
covariant derivatives on $\cM$ and $\Sigma_t$ by $\nabla$ and $D$,
respectively. The spatial covariant derivative $D$ of a $(k,l)$-tensor
field $T$ on $\Sigma_t$ is given in terms of the covariant derivative
$\nabla$ of spacetime as
\begin{equation}\label{DmuT}
D_\mu
T^{\nu_1\cdots\nu_k}_{\phantom{\nu_1\cdots\nu_k}\rho_1\cdots\rho_l} =
\projector{\sigma}{\mu}
\projector{\nu_1}{\alpha_1}\cdots\projector{\nu_k}{\alpha_k}
\projector{\beta_1}{\rho_1}\cdots\projector{\beta_l}{\rho_l}
\nabla_\sigma
T^{\alpha_1\cdots\alpha_k}_{\phantom{\alpha_1\cdots\alpha_k}
\beta_1\cdots\beta_l} \,,
\end{equation}
where in the right-hand side one considers the extension of $T$ on
spacetime. For further details on the notation, one can also see
\cite{Chaichian:2011sx}.

\subsection{Spacetime decomposition of the action}

Instead of deriving the general Arnowitt-Deser-Misner (ADM)
representation of the action \eqref{S3}, as was first done in our
analysis of the original formulation of CRG \cite{Chaichian:2011sx},
we directly consider a specially chosen foliation of spacetime.
The constraint \eqref{phiconstraint} on $\phi$ ensures that the vector
$\p^\mu \phi$ is timelike everywhere, when $U_0>0$ is assumed. Therefore
there exists a preferred foliation of spacetime into spatial
hypersurfaces $\Sigma_t$ whose unit normal is given by
\begin{equation}\label{n:pphi}
n^\mu = -\frac{\p^\mu \phi}{\sqrt{-\p_\nu \phi \p^\nu \phi}} =
-\frac{\p^\mu \phi}{\sqrt{2U_0}} \,.
\end{equation}
Then the constraint \eqref{phiconstraint} reduces to the condition that
the normal has unit norm
\begin{equation}
n_\mu n^\mu = -1 \,,
\end{equation}
which is presumed for the unit normal by definition.
Hence we are effectively substituting the solution of the equation of
motion \eqref{phiconstraint} back into the action.
The projector \eqref{scalarprojector} becomes the orthogonal projector
onto the spatial hypersurface $\Sigma_t$, $\Proj{\alpha}{\mu} =
\projector{\mu}{\alpha}$, which is defined as
\begin{equation}\label{projector}
\projector{\mu}{\alpha} = \delta_\alpha^\mu+n^\mu n_\alpha \,.
\end{equation}
Now we can write
\begin{equation}\label{pphi}
\p^\mu \phi = -\sqrt{2U_0} n^\mu \,,\qquad \p_\mu \phi = -\sqrt{2U_0}
n_\mu \,.
\end{equation}
From \eqref{normal} and \eqref{pphi} we see that in this foliation
$\phi$ is constant on $\Sigma_t$, $\phi=\phi(t)$. Clearly this
construction bears some similarity to the St\"uckelberg formalism
used in HL gravity \cite{Blas:2009,Blas:2010b}, where the foliation
structure of spacetime is encoded into a scalar field in order to
achieve a manifestly covariant action and ``transfer'' the extra
degree of freedom from the metric to the scalar. Here the scalar
field $\phi$ is present from the beginning and we choose to work
with a foliation of spacetime defined by $\phi$ via \eqref{n:pphi}.
Indeed, one could choose to decompose the action with respect to an
arbitrary foliation of spacetime, with no relation to $\phi$, but in
that case the action turns out quite complicated, involving
higher-order time derivatives up to fifth order. Note that $\phi$ is
no longer an independent variable, but rather related to $U_0$ and
the lapse $N$. The constraint \eqref{phiconstraint} on $\phi$
reduces to
\begin{equation}\label{phiconstraint.gf}
-\frac{\dot{\phi}^2}{2N^2} + U_0 = 0 \,.
\end{equation}
and we can integrate it for the scalar $\phi$
\begin{equation}\label{phi(t)}
\phi(t) = \phi(t_0) + \sqrt{2U_0} \int_{t_0}^t dt' N(t') \,.
\end{equation}
Evidently the choice $N=1/\sqrt{2U_0}$ corresponds to actually choosing
the scalar field as the time coordinate, $\phi=t$.
The relation \eqref{phiconstraint.gf} or \eqref{phi(t)} implies that the
lapse $N$ must be constant on $\Sigma_t$ too, $N=N(t)$, because
otherwise $\phi$ could not be constant on $\Sigma_t$.
In order to preserve the conditions $\phi=\phi(t)$ and $N=N(t)$ we
restrict the symmetry under diffeomorphisms of spacetime to the symmetry
under foliation-preserving diffeomorphisms \eqref{fp-diffeomorphism},
the main symmetry group of HL gravity.
In the language of Ho\v{r}ava's theory we would say that both $\phi$ and
$N$ are projectable -- like the lapse is in the projectable version of
HL gravity. Here we consider \eqref{fp-diffeomorphism} as a partial
gauge fixing of the diffeomorphism symmetry, which is required by the
choice of working with the preferred foliation.

Now the action \eqref{S3} can be written
\begin{multline}\label{S3.gf}
S_3 = \int d^4 x \sqrt{-\gM} \left[ \frac{\RM}{2\kappa^2} - \alpha
\projector{\mu}{\alpha}\projector{\nu}{\beta} \left( \RM_{\mu\nu} -
\nabla_n K_{\mu\nu} + a_\mu a_\nu \right) \right.\\
\times \left. 2U_0 \left( n^\mu n^\nu \nabla_\mu \nabla_\nu + \nabla^\mu
\nabla_\mu \right) g^{\mu\alpha}g^{\nu\beta} \left( \RM_{\mu\nu} -
\nabla_n K_{\mu\nu} + a_\mu a_\nu \right) \right] \,,
\end{multline}
where we denote $\nabla_n = n^\mu \nabla_\mu$ and use \eqref{pphi} as
$\nabla_\mu \phi = -\sqrt{2U_0}n_\mu$ and apply the following
geometrical identities
\begin{align}
\nabla_\mu n_\nu &= K_{\mu\nu} - n_\mu a_\nu \,,\\
a_\mu &= \nabla_n n_\mu = D_\mu \ln N \,,
\end{align}
in order to write
\begin{equation}
\frac{1}{2U_0}\p_\rho \phi \nabla^\rho \nabla_\mu \nabla_\nu \phi
=\nabla_n K_{\mu\nu} - a_\mu a_\nu - n_\mu \nabla_n a_\nu \,.
\end{equation}
The vector $a_\mu$ has the physical interpretation of being the
acceleration of an observer with 4-velocity $n^\mu$. It is always
orthogonal to $n^\mu$, $a_\mu n^\mu=0$, and hence tangent to
$\Sigma_t$, $\projector{\mu}{\alpha}a_\mu=a_\alpha$.

In the action \eqref{S3.gf}, we recognize
$\projector{\mu}{\alpha}\projector{\nu}{\beta}
\RM_{\mu\nu}$ as the component of the Ricci tensor of spacetime that is
tangent to $\Sigma_t$. With the help of the Gauss relation
\eqref{GaussEq} and the Ricci equation \eqref{RicciEq} it can be written
\begin{equation}\label{projRicci}
\projector{\mu}{\alpha}\projector{\nu}{\beta} \RM_{\mu\nu} =
R_{\alpha\beta} + K K_{\alpha\beta} -
2K_{\alpha\mu}K^\mu_{\phantom{\mu}\beta} - \frac{1}{N}D_\alpha D_\beta N
+\frac{1}{N}\cL_{Nn}K_{\alpha\beta}\,,
\end{equation}
where $R_{\alpha\beta}$ is the Ricci tensor of the hypersurface
$\Sigma_t$ and $\cL_{Nn}$ denotes the Lie derivative along the 4-vector
$Nn^\mu=(1,-N^i)$.
Then we calculate the Lie derivative in \eqref{projRicci}
\begin{equation}\label{LieNnK}
\frac{1}{N}\cL_{Nn}K_{\alpha\beta} = \nabla_n K_{\alpha\beta} + \left(
K_\alpha^{\phantom{\alpha}\mu} - n_\alpha a^\mu \right)  K_{\mu\beta}+
\left( K_\beta^{\phantom{\beta}\mu} - n_\beta a^\mu \right)
K_{\alpha\mu} \,.
\end{equation}
For any tensor $T$ that is tangent to $\Sigma_t$ its Lie derivative
$\cL_{Nn}T$ is also tangent to $\Sigma_t$. This follows from $\cL_{Nn}
\projector{\mu}{\alpha}=0$. Thus we can project both sides of
\eqref{LieNnK} onto $\Sigma_t$ and obtain
\begin{equation}\label{LieNnK2}
\frac{1}{N}\cL_{Nn}K_{\alpha\beta} =
\projector{\mu}{\alpha}\projector{\nu}{\beta} \nabla_n K_{\mu\nu} +
2K_{\alpha\mu}K^\mu_{\phantom{\mu}\beta} \,.
\end{equation}
Substituting \eqref{LieNnK2} into \eqref{projRicci} enables us to write
\begin{equation}\label{projRicci.2}
\projector{\mu}{\alpha}\projector{\nu}{\beta} \RM_{\mu\nu} =
R_{\alpha\beta} + K K_{\alpha\beta} - \frac{1}{N}D_\alpha D_\beta N
+ \projector{\mu}{\alpha}\projector{\nu}{\beta} \nabla_n K_{\mu\nu}
\,.
\end{equation}
Thus in the action \eqref{S3.gf} we obtain the decomposition
\begin{equation}\label{proj:Ricci-nablanK}
\projector{\mu}{\alpha}\projector{\nu}{\beta} \left( \RM_{\mu\nu} -
\nabla_n K_{\mu\nu} + a_\mu a_\nu \right)
= R_{\alpha\beta} + K K_{\alpha\beta} - D_\alpha a_\beta \,,
\end{equation}
where we can actually drop the last term in the right-hand side,
$-D_\alpha a_\beta = a_\alpha a_\beta - \frac{1}{N}D_\alpha D_\beta N$,
since the condition $N=N(t)$ implies that $a_\mu$ vanishes. In ADM
coordinates, the components of $a_\mu$ are
\begin{equation}\label{a=0}
a_0=N^ia_i=N^i\p_i\ln N=0\,,\qquad a_i=\p_i\ln N=0\,,
\end{equation}
since $\p_i N=0$. All the terms in the right-hand side of
\eqref{proj:Ricci-nablanK} are tangent to $\Sigma_t$,
which is the essential geometrical implication for introducing the
projectors into the action of new CRG.

Using \eqref{proj:Ricci-nablanK} we can write the action \eqref{S3.gf}
as
\begin{equation}\label{S3.gf.2}
S_3 = \int d^4 x \sqrt{-\gM} \left[ \frac{\RM}{2\kappa^2} - 2\alpha
U_0 \left( R^{\alpha\beta} + K K^{\alpha\beta} \right)
\left( n^\mu n^\nu \nabla_\mu \nabla_\nu + \nabla^\mu \nabla_\mu \right)
\left( R_{\alpha\beta} + K K_{\alpha\beta} \right) \right] \,.
\end{equation}
What remains to be decomposed  are the covariant derivatives of tensor
fields that are tangent to the spatial hypersurfaces $\Sigma_t$, namely
\begin{equation}\label{covdertodecompose}
\nabla_\mu \nabla_\nu \left( R_{\alpha\beta} + K K_{\alpha\beta} \right)
\,.
\end{equation}
In the original formulation of CRG one instead takes covariant
derivatives of a scalar, namely the component of the Einstein tensor
that is purely normal to $\Sigma_t$, i.e. $\p^\mu \phi \p^\nu \phi
\GM_{\mu\nu} \propto n^\mu n^\nu \GM_{\mu\nu} = \frac{1}{2}\left(
K^2-K_{ij}K^{ij}+R \right)$, which are much easier to decompose.
The relation \eqref{DmuT} of $D$ and $\nabla$ can be used in the
decomposition of covariant derivatives of tensor fields tangent to
$\Sigma_t$. In Appendix~\ref{appendix3}, we present the calculation
of spacetime decomposition of first- and second-order covariant
derivatives of a symmetric tensor field $A_{\alpha\beta}$ that is
tangent to $\Sigma_t$.

Here we consider the specific combination of second-order covariant
derivatives that appears in the action \eqref{S3.gf.2}:
\begin{equation}\label{covdersdecomp}
\begin{split}
\left( n^\mu n^\nu \nabla_\mu \nabla_\nu + \nabla^\mu \nabla_\mu \right)
A_{\alpha\beta} &=
g^{\nu\mu}\projector{\rho}{\mu}\projector{\sigma}{\alpha}\projector{
\lambda}{\beta}\nabla_\nu \nabla_\rho  A_{\sigma\lambda} - n_\alpha
g^{\nu\mu}\projector{\rho}{\mu}n^{\sigma}\projector{\lambda}{\beta}
\nabla_\nu \nabla_\rho  A_{\sigma\lambda} \\
&- n_\beta
g^{\nu\mu}\projector{\rho}{\mu}\projector{\sigma}{\alpha}n^{\lambda}
\nabla_\nu \nabla_\rho  A_{\sigma\lambda} + n_\alpha n_\beta
g^{\nu\mu}\projector{\rho}{\mu}n^{\sigma}n^{\lambda}\nabla_\nu
\nabla_\rho  A_{\sigma\lambda}\\
&= D^{\mu}D_\mu A_{\alpha\beta} - K\left(
\frac{1}{N}\cL_{Nn}A_{\alpha\beta} -
2K_{(\alpha}^{\phantom{(\alpha}\mu}A_{\beta)\mu} \right) \\
&+ 2K^{\mu\nu}K_{\mu(\alpha|}A_{\nu|\beta)} \\
&+ n_\alpha \left( 2K^{\mu\nu}D_\mu A_{\nu\beta} + D^\mu
K_{\mu\nu}A^\nu_{\phantom\nu\beta} - K a^\mu A_{\mu\beta} \right)\\
&+ n_\beta \left( 2K^{\mu\nu}D_\mu A_{\alpha\nu} + D^\mu
K_{\mu\nu}A_\alpha^{\phantom\alpha\nu} - K A_{\alpha\mu} a^\mu \right)
\\
&+ n_\alpha n_\beta 2K^{\mu\nu}K_\mu^{\phantom\mu\rho}A_{\nu\rho} \,.
\end{split}
\end{equation}
When we multiply \eqref{covdersdecomp} with another symmetric tensor
field $B^{\alpha\beta}$ that is tangent to $\Sigma_t$, we obtain
\begin{equation}\label{covdersinS}
\begin{split}
B^{\alpha\beta} \left( n^\mu n^\nu \nabla_\mu \nabla_\nu + \nabla^\mu
\nabla_\mu \right) A_{\alpha\beta} = B^{ij} &\left( D^k D_k A_{ij} - K
\frac{1}{N}\cL_{Nn}A_{ij} \right.\\
&+ 2K K_{i}^{\phantom{i}k}A_{kj} + 2K_{ik}K^{kl}A_{lj} \Bigr) \,.
\end{split}
\end{equation}
In the right-hand side of \eqref{covdersinS} all the tensor fields are
tangent to $\Sigma_t$, which enabled us to write the contractions over
spatial components.
Substituting
\begin{equation}
A_{ij} = R_{ij} + K K_{ij} \,,\qquad B^{ij} = R^{ij} + K K^{ij}
\end{equation}
into \eqref{covdersinS} gives the ADM representation of the
action \eqref{S3.gf}
\begin{multline}\label{S3.gf.ADM}
S_3 = \int d^4 x \sqrt{g}N \left\{ \frac{K_{ij}K^{ij} - K^2 +
R}{2\kappa^2}
- 2\alpha U_0 \left( R^{ij} + K K^{ij} \right)
\biggl[ D^k D_k \left( R_{ij} + K K_{ij} \right) \right.\\
-\left.\left. K \frac{1}{N}\cL_{Nn} \left( R_{ij} + K K_{ij} \right)
+ 2K K_{i}^{\phantom{i}k} \left( R_{kj} + K K_{kj} \right)
+ 2K_{ik}K^{kl} \left( R_{lj} + K K_{lj} \right) \right] \right\} \,.
\end{multline}
Here we have also included the usual decomposition of the scalar
curvature of spacetime \eqref{RM} and dropped the divergence terms; we
assume the spatial manifold is compact and that it has no boundary.
The first thing to note is that the Lagrangian of the new CRG given in
\eqref{S3.gf.ADM} contains kinetic terms involving the extrinsic
curvature to the sixth power, while conventional gravitational
Lagrangians only include kinetic terms quadratic in the extrinsic
curvature. The action also contains second-order time derivatives of
the metric in the term involving the Lie derivative of the extrinsic
curvature. Namely in the Lagrangian we have a term that contains
\begin{equation}\label{LieNnR+K2}
\begin{split}
K\left( R^{ij} + KK^{ij} \right)
\cL_{Nn} \left( R_{ij} + KK_{ij} \right) &=
K\left( R^{ij} + K K^{ij} \right)
\left[ \p_t \left( R_{ij} + KK_{ij} \right) \right.\\
&\qquad-\left. \cL_{\bm{N}} \left( R_{ij} + KK_{ij} \right) \right]  \,,
\end{split}
\end{equation}
where $\cL_{\bm{N}}$ denotes the Lie derivative along the shift vector
$N^i$ in $\Sigma_t$.
Spacetime decomposition of the Lie derivative $\cL_{Nn}$ of a covariant
tensor tangent to $\Sigma_t$ can be found in Appendix~\ref{appendix3}.

Recall that in the linearized treatment \cite{Kluson:2011rs} the part of
the action with the coupling constant $\alpha$ contained only spatial
derivatives of the perturbation to Minkowski metric. Here we obtain that
our ADM representation of the same part of the action contains first and
second-order time derivatives. Therefore already at this point we can
expect to obtain some results that will differ from
\cite{Kluson:2011rs}, in particular due to the presence of the higher
time derivative term \eqref{LieNnR+K2}.

ADM representations of CRG actions corresponding to values of the
critical exponent $z$ other than $z=3$ can be written by using the
results \eqref{proj:Ricci-nablanK} and \eqref{covdersdecomp}. It is
found that in general such actions contain kinetic terms involving the
extrinsic curvature and possibly its time derivatives to the $2z$-th
power. Higher-order time derivatives are present because every instance
of the derivative operator decomposed in \eqref{covdersdecomp} adds
one more time derivative. The only exception is the nonrenormalizable
case $z=2$, where the action is like \eqref{S3.gf.2} but without the
derivative operator decomposed in \eqref{covdersdecomp}, which involves
only first-order time derivatives albeit its kinetic part is still
modified substantially. We will not present the ADM representations of
the actions for higher $z$ here since generalization of the present
results is straightforward.

ADM representations of other conceivable generally covariant higher
derivative theories of gravity which aim to be power-counting
renormalizable by involving spontaneous (constraint induced) Lorentz
and/or diffeomorphism symmetry breaking will likely share some
characteristics with CRG. In particular, it is unlikely that all higher
time derivatives could be canceled by such construction. However,
constraints that also contain higher time derivatives might just be
able to serve such purpose.

\section{First-order Lagrangian for Hamiltonian formalism}\label{sec4}

A Hamiltonian formulation for higher derivative theories with regular
Lagrangians was first developed by Ostrogradski
\cite{Ostrogradski:1850}. Some decades after Dirac developed
Hamiltonian formalism for constrained systems
\cite{Dirac:1950,Dirac:1958} it was generalized to higher derivative
theories \cite{Gitman:1983,Buchbinder:1985,Pons:1989}. Soon
Hamiltonian formulations of higher derivative theories of gravity
were constructed for the first times
\cite{Kaku:1983,Boulware:1984,Buchbinder:1987}.
Hamiltonian formulation of actions that involve higher-order time
derivatives requires one to introduce a pair of new independent
variables for each higher-order time derivative of a variable.
There generally exist several different choices of such
additional variables, which each yield a different Hamiltonian
formulation of a given higher-order action. Such Hamiltonian
formulations are connected by canonical transformations
\cite{Buchbinder:1987} and hence they are classically equivalent.
But those canonical transformation can be highly nonlinear.
Thus there is no guarantee that the Hamiltonian formulations remain
equivalent after quantization.

Regular Lagrangians that depend on higher-order time derivatives of
dynamical variables are known to possess degrees of freedom which carry
both negative and positive energies \cite{Ostrogradski:1850}. In an
interacting higher derivative field theory such ghosts necessarily
destabilize the theory when the Lagrangian is regular (nondegenerate),
since any state of the system can and will further decay into
excitations with compensating negative and positive energies; for an
example, see \cite{Eliezer:1989}. Higher derivative theories whose
Lagrangian are singular (degenerate) can sometimes avoid the
Ostrogradskian instability. Theories which posses continuous
symmetries are always degenerate, in particular gauge theories such as
gravity, and hence they have a chance to avoid the instability.
Thus for higher derivative theories of gravity, the existence and
behavior of ghosts has to be checked in each theory. The same applies
to other pathologies such as strong coupling of extra degrees of
freedom.

A relatively simple example of a higher derivative theory of gravity is
provided by $f(\RM)$ gravity, whose Lagrangian is an arbitrary function
of the scalar curvature $\RM$ of spacetime. The scalar curvature is
second order in time derivatives \eqref{RM}. As a result the theory
contains an extra degree of freedom compared to GR, whose
Einstein-Hilbert Lagrangian is just $\RM$. For a recent review of
$f(\RM)$ gravity, see for example \cite{Sotiriou:2010}.
$f(\RM)$ gravity has several known Hamiltonian formulations based on
different choices for the higher-order variable associated with the
extra degree of freedom. One can regard the scalar curvature of
spacetime as the additional dynamical variable \cite{Boulware:1984}.
 Since the second-order time derivative in the scalar curvature
\eqref{RM} appears as $\dot{K}$, we can follow the approach of
\cite{Buchbinder:1987} by regarding the trace of the extrinsic curvature
as an alternative scalar variable. A third alternative is provided by
the fact that $f(\RM)$ gravity is equivalent to a scalar-tensor theory
where the scalar is minimally coupled to GR. The extra scalar degree of
freedom in $f(\RM)$ gravity is indeed stable if the potential is well
behaved. For a discussion of the Ostrogradskian instability and the
reason why $f(\RM)$ gravity is able to avoid it, see
\cite{Woodard:2007}. For a recent review of the aforementioned three
Hamiltonian formulations of $f(\RM)$ gravity, see \cite{Deruelle:2009}.

In the action \eqref{S3.gf.ADM}, there is a term \eqref{LieNnR+K2} that
contains a time derivative of each component of the extrinsic curvature
$K_{ij}$. That is a second-order time derivative of each component of
the metric. Therefore we shall introduce two new symmetric rank 2 tensor
variables. The way in which these new variables are defined is a matter
choice. Since our action contains the second-order time derivatives in
$\dot{K}$ and $\dot{K}_{ij}$ it would be natural to follow the approach
originated in \cite{Buchbinder:1987} where $K_{ij}$ are taken as
additional independent variables. We, however, choose alternative
variables that simplify the action further.
In order to obtain an action whose Lagrangian is quadratic in the
extrinsic curvature and the first time derivatives of the additional
higher-order variables we shall introduce an additional independent
symmetric tensor variable $\zeta_{ij}$, which is related to the metric
variables by
\begin{equation}\label{zeta_ij}
\zeta_{ij} = R_{ij} + K K_{ij}
\end{equation}
This choice also enables us to get rid of terms that involve the time
derivative $\dot{R}_{ij}$ of the spatial Ricci tensor.
In order to enforce the relation \eqref{zeta_ij} another symmetric
tensor field $\lambda^{ij}$ will be introduced as a Lagrange multiplier.
We replace the action \eqref{S3.gf.ADM} by
\begin{multline}\label{S3.gf.ADM.2}
S_3 = \int d^4 x \sqrt{g}N \left[ \frac{K_{ij}K^{ij} - K^2 +
R}{2\kappa^2}
- 2\alpha U_0 g^{ik}g^{jl} \zeta_{kl} \left( D^k D_k \zeta_{ij} -
K \frac{1}{N}\cL_{Nn} \zeta_{ij} \right.\right.\\
+\left. 2K K_{i}^{\phantom{i}k}\zeta_{kj}
+ 2K_{ik}K^{kl}\zeta_{lj} \right)
- \alpha\lambda^{ij} \left( \zeta_{ij} - R_{ij} - K K_{ij} \right)
\Bigr]\,.
\end{multline}
Variation of the action with respect to $\lambda^{ij}$ gives
Eq.~\eqref{zeta_ij}. Substituting it back into the action would give the
original action \eqref{S3.gf.ADM}.
Since the action does not contain any time derivatives of $\lambda^{ij}$
these variables have an auxiliary character in the Lagrangian. In the
action \eqref{S3.gf.ADM.2}, the term that contains a time derivative of
$\zeta_{ij}$ can be written as
\begin{equation}\label{LieNnzeta}
2\alpha U_0 Kg^{ik}g^{jl} \zeta_{kl}\cL_{Nn} \zeta_{ij} =
2\alpha U_0 Kg^{ik}g^{jl} \zeta_{kl}\left(  \dot{\zeta}_{ij} -
\cL_{\bm{N}} \zeta_{ij} \right) \,,
\end{equation}
where $\cL_{\bm{N}}$ denotes the Lie derivative along the shift vector
$N^i$ in $\Sigma_t$. Thus at least some of the variables $\zeta_{ij}$
are expected to carry propagating degrees of freedom.

In the actions \eqref{S3.gf.ADM} and \eqref{S3.gf.ADM.2} $N$
had to be restricted to a function of time only as was discussed in
Sec.~\ref{sec3}. In these actions, it would be possible to promote the
lapse function to possess dependence on both space and time, since the
symmetry under \eqref{fp-diffeomorphism} would be retained. But the
resulting action might possess very different properties compared to
the ADM representation we have here derived from the CRG action
\eqref{S3}. Hamiltonian structure of such a generalized action would
certainly bare some differences, in particular the existence of a local
Hamiltonian constraint, and its study would be interesting. Here we
shall, however, constrain ourselves to the study of the action
\eqref{S3.gf.ADM.2} with the projectability of $N$ implied by the
constraint on the scalar field \eqref{phiconstraint.gf}.

Two alternative sets of variables for Hamiltonian formulation of new CRG
are discussed in Sec.~\ref{sec7}.

\section{On solutions for the first-order Lagrangian}\label{sec5}
Let us consider solutions for the first-order Lagrangian, i.e.,
configurations for which the action \eqref{S3.gf.ADM.2} is extremal.
When the Lagrange multiplier fields $\lambda^{ij}$ are set to zero, the
equations of motion obtained by varying the higher-order variables
$\zeta_{ij}$ are linear and homogeneous in $\zeta_{ij}$. They have a
static solution $\zeta_{ij}=0$ provided that the initial value for
$\zeta_{ij}$ is zero everywhere on the $t=0$ hypersurface $\Sigma_0$.
Then the equations of motion \eqref{zeta_ij} obtained by varying
$\lambda^{ij}$ give
\begin{equation}\label{R+K2}
R_{ij}+KK_{ij}=0\,,
\end{equation}
which imply that the scalar intrinsic and extrinsic curvature are
related by $R=-K^2$. Substituting these solutions back into the
action
gives
\begin{equation}\label{S.ul}
S = \int d^4 x \sqrt{g}N\frac{K_{ij}K^{ij}-K^2+R}{2\kappa^2}
= \int d^4 x \sqrt{g}N\frac{K_{ij}K^{ij}-2K^2}{2\kappa^2}
\end{equation}
with the intrinsic and extrinsic curvature related by \eqref{R+K2}.
This kind of action with no curvature dependent terms in the potential
is called ultralocal gravity. Such an ultralocal theory of gravity where
the kinetic part of the action is identical to that of GR and the
potential consists only of the cosmological constant was originally
proposed in \cite{Teitelboim:1980}. The Hamiltonian of ultralocal
gravity has the same form as in GR, written in terms of the ADM
variables as
\begin{equation}
H=\int d^3\bx\left( N\cH_0+N^i\cH_i \right)\,,
\end{equation}
where $\cH_0$ is the Hamiltonian constraint and $\cH_i$ is the
momentum constraint. The appealing characteristic of ultralocal gravity
is that the algebra of constraints is a true Lie algebra, because  the
Poisson bracket of Hamiltonian constraints $\cH_0(\bx)$ and
$\cH_0(\by)$ is zero. In GR the Poisson bracket of Hamiltonian
constraints is equal to a linear combination of momentum constraints
$\cH_i(\bx)$ with field-dependent ``structure constants''. Thus the
constraints of GR lack a Lie algebra structure. The momentum constraint
of ultralocal gravity is identical to that of GR and hence the rest of
the constraint algebra of ultralocal gravity coincides with GR.

Note that the kinetic part of the action \eqref{S.ul} is not identical
to GR when we take into account the relation \eqref{R+K2} as $R=-K^2$,
since the $K^2$ term is modified by the factor 2. Alternatively we could
obtain both a modified kinetic part and a modified potential:
$K_{ij}K^{ij}-K^2+R=K_{ij}K^{ij}+2R$. Rather the action \eqref{S.ul}
corresponds to the ultralocal case of HL gravity
\cite{Horava:2009uw} in the limit $z\rightarrow0$, $\lambda\rightarrow2$
and with vanishing cosmological constant. Here $\lambda$ is the
coupling modifying the kinetic part of the Lagrangian:
$K_{ij}K^{ij}-\lambda K^2$. The crucial difference is that in this
class of solutions the spatial Ricci tensor and the extrinsic curvature
are always related by \eqref{R+K2}.
A special case of this kind of solution is the static flat spacetime for
which $R_{ij}=0$ and $K_{ij}=0$ satisfy \eqref{R+K2}. Thus we confirm
that CRG action has the static flat solution. However, this does not
necessarily mean that it is a stable vacuum state.
Most physically interesting solutions do not respect the relation
\eqref{R+K2} and hence we must look for more general dynamics.

When the initial geometry of spacetime deviates from \eqref{R+K2} even
slightly, the dynamics of spacetime produced by the action
\eqref{S3.gf.ADM.2} is very different compared to the class of solutions
discussed above. The equations of motion obtained by varying $g^{ij}$
and $\zeta_{ij}$ are highly nontrivial. Indeed the dynamics of the
metric and the higher-order variables $\zeta_{ij}$ is very complicated.
Thus we do not attempt to solve the equations of motion in the general
case. Instead we seek to understand the system through Hamiltonian
analysis.

\section{Hamiltonian analysis}\label{sec6}

Let us analyze the new CRG action \eqref{S3.gf.ADM.2} by using the
Hamiltonian formalism generalized for constrained systems
\cite{Dirac:1950,Dirac:1958}. For reviews, see
\cite{HamiltonianAnalysis}.

In Sec.~\ref{sec5}, we saw that when $\zeta_{ij}=0$, i.e., when
\eqref{R+K2} holds, the action reduces to an ultralocal special case of
HL gravity \eqref{S.ul}. Now we aim to understand the structure and
dynamics of the theory when $\zeta_{ij}$ is nonzero. We assume that at
least two components of $\zeta_{ij}$ are nonzero, while the rest of the
components can attain any values. In principle it makes no difference
which of the components are chosen to be nonzero.
We choose one of the nonvanishing components to be the
trace of $\zeta_{ij}$ due to notational elegance.

First we define the canonical momenta. Since the action
\eqref{S3.gf.ADM.2} is independent of the time derivatives of $N$,
$N^i$, and $\lambda^{ij}$, their canonically conjugated momenta, $p_N$,
$p_i$, and $p^{\lambda}_{ij}$, respectively, are primary constraints:
\begin{equation}\label{pconstraints}
p_N \approx 0 \,,\qquad p_i(\bx) \approx 0 \,,\qquad
p^{\lambda}_{ij}(\bx) \approx 0 \,.
\end{equation}
The momenta canonically conjugate to $g_{ij}$ and $\zeta_{ij}$ are
defined by
\begin{align}
p^{ij} = \frac{\delta S_3}{\delta\dot{g}_{ij}} &= \sqrt{g}\left[
\frac{K^{ij}-g^{ij}K}{2\kappa^2} + \frac{\alpha U_0}{N}g^{ij} \left(
\dot{\zeta}_{kl}-\cL_{\bm{N}}\zeta_{kl} \right) g^{km}g^{ln}\zeta_{mn}
\right.\nn\\
&\quad- 2\alpha U_0 \left( g^{ij}K^{kl} + g^{ik}K^{jl} + g^{jk} K^{il}
+ g^{ik}g^{jl}K \right) \zeta_{km}\zeta_{ln}g^{mn} \nn\\
&\quad+\left. \frac{\alpha}{2}\left( g^{ij}\lambda^{kl}K_{kl} +
\lambda^{ij}K
\right)
\right] ,\label{p^ij}\\
p_{\zeta}^{ij} = \frac{\delta S_3}{\delta\dot{\zeta}_{ij}}
&= \sqrt{g}2\alpha U_0 K g^{ik}g^{jl}\zeta_{kl} \,.\label{p_zeta^ij}
\end{align}
We shall adopt a convention where the trace component of a tensor or a
tensor density is denoted without indices and the traceless component is
denoted with the bar accent. For instance according to this convention
we decompose the variables $\zeta_{ij}$ and $p_{\zeta}^{ij}$ as
\begin{equation}
\zeta_{ij}=\bar{\zeta}_{ij}+\frac{1}{3}g_{ij}\zeta\,,\qquad
p_{\zeta}^{ij}=\bar{p}_{\zeta}^{ij}+\frac{1}{3}g_{ij}p_\zeta\,,
\end{equation}
where $\bar{\zeta}_{ij}$ and $\bar{p}_{\zeta}^{ij}$ are the traceless
components and $\zeta=g^{ij}\zeta_{ij}$ and
$p_{\zeta}=g_{ij}p_{\zeta}^{ij}$ are the trace components. The same
notation can be used for the other variables and tensors in general.
Taking the traces of the momenta \eqref{p^ij} and \eqref{p_zeta^ij}
gives
\begin{align}
p &= \sqrt{g}\left[ -\frac{K}{\kappa^2}
+ \frac{3\alpha U_0}{N}\left( \dot{\zeta}_{ij}-\cL_{\bm{N}}
\zeta_{ij} \right) g^{ik}g^{jl}\zeta_{kl}
- \alpha U_0 \left(10K^{ij}+2Kg^{ij}\right)\zeta_{ik}\zeta_{jl}g^{kl}
\right.\nn\\
&\qquad\quad+ \frac{\alpha}{2}\left( 3\lambda^{ij}K_{ij} + \lambda K
\right)
\Bigr] \,,\label{p.z3}\\
p_{\zeta} &= \sqrt{g}2\alpha U_0 K\zeta \,.\label{p_zeta}
\end{align}
where we use the aforementioned convention to denote
$p=g_{ij}p^{ij}$, $\lambda=g_{ij}\lambda^{ij}$ etc.
Assuming $\zeta_{ij}$ has a nonvanishing trace, $\zeta\neq0$, and the
coupling constants $\alpha\neq0$ and $U_0>0$,
we solve \eqref{p_zeta} for the trace of the extrinsic curvature
\begin{equation}\label{K.solved}
K = \frac{1}{\sqrt{g}}\frac{p_{\zeta}}{2\alpha U_0 \zeta}
\end{equation}
Additional primary constraints can be obtained by substituting
\eqref{K.solved} back into \eqref{p_zeta^ij}. The trace of the resulting
equation is a trivial identity, but its traceless part expresses
$\bar{p}_{\zeta}^{ij}$ in terms of other variables.
Thus we obtain five new primary constraints
\begin{equation}\label{barPi}
\bar{\Pi}^{ij} = \bar{p}_{\zeta}^{ij}-
g^{ik}g^{jl}\bar{\zeta}_{kl}\frac{p_\zeta}{\zeta} \approx 0\,,
\end{equation}
which form a symmetric traceless tensor density. This primary constraint
tells us that only the trace component of the momenta $p_{\zeta}^{ij}$
is an independent variable.
In order to solve the traceless component $\bar{K}_{ij}$ of the
extrinsic curvature we separate it from the already solved trace
component \eqref{K.solved} as
\begin{equation}\label{K_ij.decom}
K_{ij} = \bar{K}_{ij} + \frac{1}{3}g_{ij}K = \bar{K}_{ij} +
\frac{1}{\sqrt{g}}\frac{g_{ij}}{6\alpha U_0}\frac{p_{\zeta}}{\zeta} \,.
\end{equation}
Then we substitute \eqref{K.solved} and \eqref{K_ij.decom} into
\eqref{p^ij} and \eqref{p.z3}. We obtain the equation
\begin{equation}\label{solve:barK_ij}
\bar{P}^{ij} = \frac{\sqrt{g}}{2\kappa^2}F^{ijkl}\bar{K}_{kl} \,,
\end{equation}
where we have defined
\begin{equation}\label{barP}
\begin{split}
\bar{P}^{ij} &= p^{ij} - \frac{1}{3}g^{ij}p + \frac{5}{3}\left(
g^{ik}g^{jl}-\frac{1}{3}g^{ij}g^{kl} \right) \zeta_{km}\zeta_{ln}g^{mn}
\frac{p_{\zeta}}{\zeta}
- \frac{1}{4U_0}\left( \lambda^{ij}-\frac{1}{3}g^{ij}\lambda \right)
\frac{p_{\zeta}}{\zeta}\\
&= \bar{p}^{ij} + \frac{5}{3}\left( \zeta^i_{\phantom{i}k}\zeta^{jk}
-\frac{1}{3}g^{ij}\zeta_{kl}\zeta^{kl} \right) \frac{p_{\zeta}}{\zeta}
- \frac{1}{4U_0}\bar{\lambda}^{ij}\frac{p_{\zeta}}{\zeta} \,,
\end{split}
\end{equation}
and
\begin{equation}\label{F^ijkl}
F^{ijkl} = \frac{1}{2}\left( g^{ik}g^{jl}+g^{il}g^{jk} \right)
- 8\kappa^2\alpha U_0 \left( g^{m(i}g^{j)(k}g^{l)n}
- \frac{1}{3}g^{ij}g^{km}g^{ln} \right) \zeta_{mo}\zeta_{np}g^{op} \,.
\end{equation}
for the purpose of shortening the following expressions for the
Hamiltonian and its constraints, which are quite complicated. 
From now on we may raise and lower spatial indices with the spatial
metric, e.g., $\zeta^i_{\phantom{i}k}=g^{ij}\zeta_{jk}$ and
$\zeta^{ij}=g^{ik}g^{jl}\zeta_{kl}$, but always keeping in mind that
the metric has to be written explicitly when Poisson bracket is
evaluated, similarly as it has to be done with the trace components,
e.g., $\zeta=g^{ij}\zeta_{ij}$.
Using the inverse $F^{-1}_{ijkl}$ to $F^{ijkl}$, such that
$F^{-1}_{ijkl}F^{klmn}=\delta_{i}^{(m}\delta_{j}^{n)}$, enables us to
solve $\bar{K}_{ij}$ from \eqref{solve:barK_ij} in terms of the
canonical variables as
\begin{equation}\label{barK_ij.solved}
\bar{K}_{ij} = \frac{2\kappa^2}{\sqrt{g}}F^{-1}_{ijkl}\bar{P}^{kl}\,.
\end{equation}
Together \eqref{K_ij.decom} and \eqref{barK_ij.solved} express the
extrinsic curvature \eqref{K_ij} in terms of the canonical variables.
The existence of $F^{-1}_{ijkl}$ means that no more primary constraints
are required. Similarly as $F^{ijkl}$, $F^{-1}_{ijkl}$ depends on the
variables $g_{ij}$ and $\zeta_{ij}$ and on the (coupling) constants. The
explicit form of $F^{-1}_{ijkl}$ will not be used in the following
analysis. Construction of $F^{-1}_{ijkl}$ can be found in
Appendix~\ref{appendix4}.

Then we shall introduce the Hamiltonian.
We define the total Hamiltonian as
\begin{equation}\label{H}
\begin{split}
H &= \int d^3\bx \left( p^{ij}\dot{g}_{ij}
+ p_{\zeta}^{ij}\dot{\zeta}_{ij} + u_N p_N + u^i p_i
+ u_\lambda^{ij}p^\lambda_{ij} + \bar{v}_{ij}\bar{\Pi}^{ij}
 - \cL \right) \\
&= \int d^3\bx \left( N\cH_0 + N^i\cH_i + u_N p_N
+ u^i p_i + u_\lambda^{ij}p^\lambda_{ij}
+ \bar{v}_{ij}\bar{\Pi}^{ij} \right) \,,
\end{split}
\end{equation}
where $u_N$, $u^i$, $u_\lambda^{ij}$ and
$\bar{v}_{ij}$ are Lagrange multipliers and we have defined
\begin{equation}\label{cH_0}
\begin{split}
\cH_0 &= \frac{1}{\sqrt{g}}\left[ 4\kappa^2p^{ij}
F^{-1}_{ijkl}\bar{P}^{kl}
+ \frac{p}{3\alpha U_0}\frac{p_\zeta}{\zeta}
- 2\kappa^2 F^{-1}_{ijkl}\bar{P}^{kl}g^{im}g^{jn}
F^{-1}_{mnop}\bar{P}^{op}
\right.\\
&\quad+
\frac{1}{12\kappa^2\alpha^2U_0^2}\left(\frac{p_\zeta}{\zeta}\right)^2
+ \frac{4}{9\alpha U_0}\zeta_{ij}\zeta^{ij}
\left(\frac{p_\zeta}{\zeta}\right)^2\\
&\quad+ \frac{20\kappa^2}{3}\zeta^{ij}\zeta^k_{\phantom{k}j}
F^{-1}_{iklm}\bar{P}^{lm}\frac{p_\zeta}{\zeta}
+ 16\kappa^4\alpha U_0 \zeta^{ij}
F^{-1}_{iklm}\bar{P}^{lm}\zeta^n_{\phantom{n}j}g^{ko}
F^{-1}_{nopq}\bar{P}^{pq} \\
&\quad-\left. \frac{\kappa^2}{U_0}\lambda^{ij}
F^{-1}_{ijkl}\bar{P}^{kl}\frac{p_\zeta}{\zeta}
- \frac{\lambda}{12\alpha U_0^2}\left(\frac{p_\zeta}{\zeta}\right)^2
 \right] \\
&\quad- \sqrt{g}\left[ \frac{R}{2\kappa^2}
- 2\alpha U_0 \zeta^{ij}D^kD_k\zeta_{ij}
- \alpha\lambda^{ij}\left( \zeta_{ij}-R_{ij} \right)
\right]
\end{split}
\end{equation}
and
\begin{equation}\label{cH_i}
\begin{split}
\cH_i &= -2g_{ij}D_k p^{jk}
+ \p_i \zeta_{jk}p_{\zeta}^{jk}
- 2\p_j \left( \zeta_{ik}p_\zeta^{jk} \right) \\
&= -2g_{ij}\p_k p^{jk} - \left( 2\p_j g_{ik} - \p_ig_{jk} \right) p^{jk}
- 2\zeta_{ij}\p_k p_\zeta^{jk}
- \left( 2\p_j\zeta_{ik} - \p_i \zeta_{jk} \right) p_{\zeta}^{jk} \,.
\end{split}
\end{equation}
Observe that the Hamiltonian only depends on the traceless components
$\bar{p}_\zeta^{ij}$ of the momenta canonically conjugate to
$\zeta_{ij}$ through the primary constraint \eqref{barPi}, which also
means that $\bar{p}_\zeta^{ij}$ does not appear in any other
constraint.

The primary constraints \eqref{pconstraints} and \eqref{barPi} must
be preserved under time evolution generated by the total Hamiltonian
\eqref{H}. For this purpose we introduce the secondary constraints
\begin{align}
&\Phi_0 = \int d^3\bx\cH_0 \approx 0\,,\label{Phi_0}\\
&\cH_i(\bx) \approx 0\,,\\
&\Psi_{ij}(\bx) = \pb{p^\lambda_{ij}(\bx),\Phi_0} \approx
0\,,\label{Psi_ij.def}\,,\\
&\bar{\Pi}_{II}^{ij}(\bx) = \pb{\bar{\Pi}^{ij}(\bx),\Phi_0} \approx
0\,,\label{barPi_II}
\end{align}
which ensure that the primary constraints $p_N$, $p_i$, $p^\lambda_{ij}$
and $\bar{\Pi}^{ij}$ are preserved in time, respectively.
The constraint \eqref{Psi_ij.def} can be explicitly defined as the
following symmetric tensor density:
\begin{equation}\label{Psi_ij}
\begin{split}
\Psi_{ij} &= \frac{1}{\sqrt{g}}\frac{\kappa^2}{U_0}
\left( \psi_{ij} - \frac{1}{3}g_{ij}g^{kl}\psi_{kl}
+ F^{-1}_{ijkl}\bar{P}^{kl} \right) \frac{p_\zeta}{\zeta}\\
&\quad+ \frac{1}{\sqrt{g}}\frac{g_{ij}}{12\alpha U_0^2}\left(
\frac{p_\zeta}{\zeta} \right)^2
- \sqrt{g}\alpha\left( \zeta_{ij}-R_{ij}
\right)
\approx 0\,,
\end{split}
\end{equation}
where we denote
\begin{equation}
\begin{split}
\psi_{ij} &= p^{kl}F^{-1}_{klij} - F^{-1}_{klij}g^{km}g^{ln}
F^{-1}_{mnop}\bar{P}^{op}
+ \frac{5}{3}\zeta^{kl}\zeta^m_{\phantom{m}l}F^{-1}_{kmij}
\frac{p_\zeta}{\zeta}\\
&\quad+ 8\kappa^2\alpha
U_0\zeta^{kl}F^{-1}_{kmij}\zeta^n_{\phantom{n}l}g^{mo}
F^{-1}_{nopq}\bar{P}^{pq}
- \frac{1}{4U_0}\lambda^{kl}F^{-1}_{klij}\frac{p_\zeta}{\zeta}\,.
\end{split}
\end{equation}
The trace of the constraint $\Psi_{ij}$ is independent of
$\lambda^{ij}$.
Indeed it can be written as
\begin{equation}\label{Psi}
\Psi = \frac{1}{\sqrt{g}}\frac{1}{4\alpha U_0^2}\left(
\frac{p_\zeta}{\zeta} \right)^2
- \sqrt{g}\alpha(\zeta-R)\,.
\end{equation}
This implies that
\begin{equation}\label{pb:Psi,p^xi_kl}
\pb{\Psi(\bx),p^\lambda_{kl}(\by)}
=g^{ij}(\bx)\pb{\Psi_{ij}(\bx),p^\lambda_{kl}(\by)}
=0\,.
\end{equation}
The traceless component of the constraint $\Psi_{ij}$ depends on the
traceless component of $\lambda^{ij}$, but it is independent of the
trace
component $\lambda$.
Therefore we shall also decompose each of the constraints
$p^\lambda_{ij}$
and $\Psi_{ij}$ into a traceless component and a trace component:
\begin{align}
p^\lambda_{ij}&=\bar{p}^\lambda_{ij}+\frac{1}{3}g_{ij}p^\lambda\,,\nn\\
\Psi_{ij}&=\bar{\Psi}_{ij}+\frac{1}{3}g_{ij}\Psi\,.
\end{align}
The explicit form of the constraint $\bar{\Pi}_{II}^{ij}$ defined in
\eqref{barPi_II} is a very complicated symmetric traceless tensor
density. We shall not write it down here since its explicit form will
not be used in our analysis. Instead when we evaluate Poisson brackets
between $\bar{\Pi}_{II}^{ij}$ and the other constraints we shall take
advantage of its definition as the Poisson bracket of $\bar{\Pi}^{ij}$
and $\Phi_0$.

Then we must ensure that every secondary constraint is preserved in
time. For the constraints $\Phi_0$ and $\cH_i$ this is easy to do, but
for the constraints $\Psi_{ij}$ and $\bar{\Pi}_{II}^{ij}$ the process
is involved and technical. The details are given in
Appendix~\ref{appendix5}.

The Hamiltonian that ensures the consistency of all the constraints in
time is
\begin{equation}\label{H.final}
H=N\Phi_0+N\int d^3\bx\left( \bar{w}_\lambda^{ij}\bar{p}^\lambda_{ij}
+\frac{1}{3}w p^\lambda+\bar{w}_{ij}\bar{\Pi}^{ij} \right)
+\int d^3\bx N^i\cH_i + v_N p_N +\int d^3\bx v^i p_i \,,
\end{equation}
where all the Lagrange multipliers denoted by $w$ have been solved in
terms of the canonical variables, while the multipliers denoted by $v$
are arbitrary. More specifically $N\bar{w}_\lambda^{ij}$
is the traceless specific solution \eqref{barw_xi} to the inhomogeneous
equation \eqref{eq.v_xi} and $Nw$ and $N\bar{w}_{ij}$ are
the specific solution to the inhomogeneous equations \eqref{pb:Psi,H}
and \eqref{eq.restofLM}.
Since the Hamiltonian is always a first-class quantity and
\eqref{H.final} is a sum of constraints, we can conclude that the
Hamiltonian is a sum of first-class constraints. That is exactly what
one expects from a system with symmetry under foliation-preserving
diffeomorphisms \eqref{fp-diffeomorphism}. Whenever a system is
invariant under time reparametrization, its Hamiltonian is a first-class
constraint. The same is true for theories with more general
diffeomorphism invariance, such as GR and the present theory. In a
generally covariant system, time evolution is just the unfolding of a
gauge transformation. We already knew that the momentum constraint
$\cH_i$, or rather its generalization $\Phi_i$ in \eqref{Phi_i}, and the
primary constraints $p_N$ and $p_i$ are first-class constraints. Now we
can see that the linear combination
\begin{equation}\label{Phi_0+.fs}
\int d^3\bx\left(\cH_0+\bar{w}_\lambda^{ij}\bar{p}^\lambda_{ij}
+\frac{1}{3}w p^\lambda+\bar{w}_{ij}\bar{\Pi}^{ij} \right)
\end{equation}
is a first-class constraint. All these first-class constraints are
associated with the symmetry under foliation preserving diffeomorphism:
invariance under time-dependent spatial diffeomorphism and time
reparametrization.

The nature of the rest of the secondary constraints remains to be
elaborated. Namely we are interested in whether some linear combination
of the secondary constraints is a first-class constraint. That is in
addition to the known first-class combination \eqref{Phi_0+.fs}, which
is indeed the only one that can include $\Phi_0$. Under the Hamiltonian
\eqref{H.2nd} we remark that including $\Psi_{ij}$ and
$\bar{\Pi}_{II}^{ij}$ in the Hamiltonian with Lagrange multipliers
would result to the fixation of the Lagrange multipliers due to the
consistency conditions of the primary constraints $p^\lambda_{ij}$ and
$\bar{\Pi}^{ij}$. This implies that every linear combination of these
constraints is a second-class constraint. Consequently there is no need
to extend the Hamiltonian \eqref{H.final} with further secondary
constraints.

Finally we can seek to identify the physical degrees of freedom. First
let us count the number of propagating physical degrees of freedom
(physical d.o.f.) by using Dirac's formula:
\begin{multline}
\#(\text{physical d.o.f.}) = \frac{1}{2}\left[ \#(\text{canonical
variables}) - 2\times\#(\text{first-class constraints}) \right.\\
\left.- \#(\text{second-class constraints}) \right] \,.
\end{multline}
We have 42 $\bx$-dependent canonical variables
($N^i$, $g_{ij}$, $\zeta_{ij}$, $\lambda^{ij}$ and
their conjugated momenta), six first-class constraints
($p_i$, $\Phi_i$) and 22 second-class constraints
($p^\lambda_{ij}$, $\bar{\Pi}^{ij}$, $\Psi_{ij}$,
$\bar{\Pi}_{II}^{ij}$):
\begin{equation}
\#(\text{physical d.o.f.}) = \frac{42-12-22}{2} = 4 \,.
\end{equation}
That is 2 more physical degrees of freedom than in GR.
There exist two additional $\bx$-independent variables $N,p_N$ and two
additional global first-class constraints $p_N$ and \eqref{Phi_0+.fs},
thus yielding 3 physical (nonpropagating) $\bx$-independent degrees of
freedom. Let us further seek to understand the nature of the extra
propagating modes in the theory.

The constraints of the theory enable us to regard some of the canonical
variables as being dependent on other variables. By analyzing the
dependencies of the variables we seek to identify the physical degrees
of freedom. First let us consider the second-class constraints.
We assume that any conceivable global first-class constraint associated
with a solution to the homogeneous part of \eqref{eq.restofLM} has been
dealt with by introducing a global gauge fixing condition, so that no
first-class constraint is set to zero strongly when we impose the
second-class constraints to zero locally, and hence the Dirac bracket
will be well defined. The Dirac bracket can be defined in the standard
way. Then we can set the local second-class constraints
$p^\lambda_{ij}$, $\bar{\Pi}^{ij}$, $\Psi_{ij}$ and
$\bar{\Pi}_{II}^{ij}$ to zero. Let us list the variables that are
regarded as dependent variables:
\begin{itemize}
\item The momenta $p^\lambda_{ij}$ canonically conjugate to
$\lambda^{ij}$ can
be set to zero. However, they did not appear in the Hamiltonian in the
first place.
\item The traceless component of $\lambda^{ij}$ can be solved from
$\bar{\Psi}_{ij}$.
\item $\Psi$ can be regarded to constrain the metric $g_{ij}$. For
instance we can regard that $\Psi$ fixes the conformal factor of the
metric $g_{ij}$. Note that we cannot use $\Psi$ to solve $\lambda$
because
$\Psi$ is independent of $\lambda^{ij}$.
\item The constraints $\bar{\Pi}^{ij}$ and $\bar{\Pi}_{II}^{ij}$
can be regarded to define the traceless variables $\bar{\zeta}_{ij}$ and
$\bar{p}_\zeta^{ij}$ in terms of independent variables.
\end{itemize}
Now the remaining independent canonical variables are 11 variables in
$g_{ij}$ and $p^{ij}$, the trace component $\lambda$, and the trace
components $\zeta$ and $p^\zeta$. We could not remove $\lambda$ from
the set of independent variables even though the variables
$\lambda^{ij}$ have an auxiliary role in the action \eqref{S3.gf.ADM.2}.

Recall that in GR the canonical variables $g_{ij}$ and $p^{ij}$ are
restricted by the Hamiltonian constraint and the momentum constraint so
that we can freely specify Cauchy data for the trace and
transverse-traceless parts of the momentum $p^{ij}$ and also the spatial
metric $g_{ij}$ up to a conformal factor
\cite{York:1971,York:1974a,York:1974b}. Then the longitudinal (vector)
part of the momentum $p^{ij}$ and the conformal factor of metric are
fixed by the momentum constraint and the Hamiltonian constraint,
respectively. Fixing the trace $p$ of the momentum fixes the slicing of
spacetime into spacelike hypersurfaces
\cite{York:1971,Dirac:1959,DeWitt:1967};
the scalar quantity $g^{-1/2}p$ measures the rate of contraction
of local three-volume elements with respect to local proper time. This
leaves us the two well-known gravitational degrees of freedom which can
be described by a pair of symmetric transverse traceless tensors.

The remaining degeneracy left in the system is due to the
time-dependent spatial diffeomorphism generated by the momentum
constraint and the invariance under time reparametrization.
Imposing the momentum constraint removes 3 degrees of freedom. Fixing
the time reparametrization symmetry does not affect local dynamics.
Thus the spatial metric tensor together with $\lambda$ carries 3
physical degrees of freedom.
The fact that the local Hamiltonian constraint is absent in
the given theory implies that there should exist an additional scalar
mode as in the projectable version of HL gravity.
But recall that here the projectability of $N$ appeared because
we chose to work with a foliation adapted to the field $\phi$
so that the spatial hypersurfaces are those of constant $\phi$.
By solving the equation of motion \eqref{phiconstraint} we have
effectively chosen the direction of the time coordinate.
Using an arbitrary spacelike foliation would not imply the
projectability of $N$. Then of course there would appear
higher derivatives of $\phi$ in the action. In that case the
extra mode would presumably be associated with $\phi$ itself.
In addition there exists another extra degree of freedom
in the trace component $\zeta$ of the tensor field $\zeta_{ij}$.
This extra mode is clearly associated with the second-order time
derivatives present in the action \eqref{S3.gf.ADM}.

In the gauge $N=1, N^i=0$, the Hamiltonian in reduced phase space is
\begin{equation}\label{H.reduced}
H=\Phi_0=\int d^3\bx\cH_0 \,,
\end{equation}
where $\cH_0$ is expressed in terms of the independent
variables in $g_{ij}$, $p^{ij}$, $\lambda$ and $\zeta$, $p_\zeta$.
However, we shall soon see that the Hamiltonian can be written without
dependence on $\lambda$, due to the constraints.

The fact that a given higher derivative theory of gravity possesses
extra degrees of freedom is always alarming, because such modes are
often ghosts or otherwise pathological. Therefore it should be checked
whether the extra modes present in this new version of CRG are
ghosts, particularly the propagating mode $\zeta$ of the higher-order
variable. We can see that the kinetic part of the Hamiltonian contains
terms involving the momentum $p_\zeta$ which can attain arbitrarily
negative values on the constraint surface. Note that this is not quite
as evident as it would seem at first sight because of the existence of
complicated constraints. Indeed a better view of this point is achieved
by taking out a linear combination of the constraints $\Psi_{ij}$
from the Hamiltonian:
\begin{equation}\label{cH_0.2}
\begin{split}
\cH_0+\lambda^{ij}\Psi_{ij} &= \frac{1}{\sqrt{g}}\left[ 4\kappa^2p^{ij}
F^{-1}_{ijkl}\bar{\cP}^{kl}
+ \frac{p}{3\alpha U_0}\frac{p_\zeta}{\zeta}
- 2\kappa^2 F^{-1}_{ijkl}\bar{\cP}^{kl}g^{im}g^{jn}
F^{-1}_{mnop}\bar{\cP}^{op} \right.\\
&\quad+ \frac{\kappa^2}{8U_0^2}F^{-1}_{ijkl}\bar{\lambda}^{kl}
g^{im}g^{jn}F^{-1}_{mnop}\bar{\lambda}^{op}
\left(\frac{p_\zeta}{\zeta}\right)^2
+\frac{1}{12\kappa^2\alpha^2U_0^2}\left(\frac{p_\zeta}{\zeta}\right)^2\\
&\quad+ \frac{4}{9\alpha U_0}\zeta_{ij}\zeta^{ij}
\left(\frac{p_\zeta}{\zeta}\right)^2
+\frac{20\kappa^2}{3}\zeta^{ij}\zeta^k_{\phantom{k}j}
F^{-1}_{iklm}\bar{\cP}^{lm}\frac{p_\zeta}{\zeta} \\
&\quad+ 16\kappa^4\alpha U_0 \zeta^{ij}
F^{-1}_{iklm}\bar{\cP}^{lm}\zeta^n_{\phantom{n}j}g^{ko}
F^{-1}_{nopq}\bar{\cP}^{pq}\\
&\quad-\left. \frac{\kappa^4\alpha}{U_0}
F^{-1}_{iklm}\bar{\lambda}^{lm}\zeta^n_{\phantom{n}j}g^{ko}
F^{-1}_{nopq}\bar{\lambda}^{pq}\left(\frac{p_\zeta}{\zeta}\right)^2
- \frac{\kappa^2}{U_0}\bar{\lambda}^{ij}
F^{-1}_{ijkl}\bar{\lambda}^{kl}\left(\frac{p_\zeta}{\zeta}\right)^2
 \right] \\
&\quad- \sqrt{g}\left( \frac{R}{2\kappa^2}
- 2\alpha U_0 \zeta^{ij}D^kD_k\zeta_{ij} \right)
\end{split}
\end{equation}
where we denote
\begin{equation}
\bar{\cP}^{kl} = \bar{p}^{kl} + \frac{5}{3}\left(
\zeta^k_{\phantom{k}m}\zeta^{lm}
-\frac{1}{3}g^{kl}\zeta_{mn}\zeta^{mn} \right)
\frac{p_{\zeta}}{\zeta}\,.
\end{equation}
This suggests that there exists a degree of freedom that carries
negative energy. As is usual in higher time derivative theories, energy
of the higher-order degree of freedom has opposite sign compared to the
corresponding lower-order degree of freedom. Here these gravitational
degrees of freedom evidently interact with each other.
We emphasize that the Hamiltonian is not the energy of the gravitational
system, but rather a constraint that vanishes everywhere on the
constraint surface. The ADM energy of an asymptotically flat
gravitational system can be obtained by studying the surface terms of
the Hamiltonian with appropriate boundary conditions.\footnote{We do
not study the total gravitational energy in this work, but rather
concentrate on the propagating degrees of freedom.}
The problem is not that the Hamiltonian or the total energy would
attain arbitrarily negative values. The real source of the problem is
that any state can decay into compensating positive and negative energy
excitations. Even ``empty space'' will decay into a tempest of positive
and negative energy excitations. This makes the theory unstable. The
only way this could be avoided are the constraints. Unfortunately the
Hamiltonian constraint is global and therefore it does not affect local
physics. We believe the momentum constraint does not protect the
stability either, since the extra degree of freedom cannot be removed
by a spatial diffeomorphism.

We note that the Hamiltonian formulation of CRG possesses a
singularity at $\zeta_{ij}=0$. In the subspace $\zeta_{ij}=0$
of phase space, all the momenta $p_\zeta^{ij}$ are primary constraints
and therefore there are no degrees of freedom. Thus there is a kind of
discontinuity between the solutions discussed in Sec.~\ref{sec5} and
the ones with nonvanishing $\zeta_{ij}$.

Some characteristics of the CRG action \eqref{S3.gf.ADM.2} can be
demonstrated by the following simple Lagrangian with two degrees of
freedom $q^a(t)$ ($a=1,2$):
\begin{equation}
L(q^a,\dot{q}^a)=\frac{1}{2}A_{ab}\dot{q}^a\dot{q}^b-V(q^a)
\,,\qquad
A=\begin{pmatrix}1 & \alpha q^2\\ \alpha q^2 & 0\end{pmatrix}\,,
\end{equation}
where $\alpha$ is a nonvanishing constant.
The canonically conjugated momenta are defined as
\begin{equation}
p_a=\frac{\p L}{\p\dot{q}^a}=A_{ab}\dot{q}^b \,.
\end{equation}
When $q^2\neq0$, $A$ has an inverse and we obtain the Hamiltonian
\begin{equation}
H(q^a,p_a)=\frac{p_1p_2}{\alpha q^2}
-\frac{1}{2}\left(\frac{p_2}{\alpha q^2}\right)^2
+V(q^a)\,.
\end{equation}
In the limit $q^2\rightarrow0$, the kinetic part of the Hamiltonian
diverges. Indeed when $q^2=0$, the conjugated momentum $p_2=0$ and the
model reduces to a single degree of freedom $q^1$. This is analogous to
the case $\zeta_{ij}=0$ of CRG.
When $q^2\neq0$, equations of motion are
\begin{align}
\dot{q}^1&=\frac{p_2}{\alpha q^2}\,,& \dot{p}_1&=-\frac{\p V}{\p
q_1}\,,\nn\\
\dot{q}^2&=\frac{p_1}{\alpha q^2}-\frac{p_2}{(\alpha q^2)^2}\,,&
\dot{p}_2&=\frac{p_1p_2}{\alpha (q^2)^2}-\frac{p_2^2}{\alpha^2
(q^2)^3}-\frac{\p V}{\p q_2}\,.
\end{align}
The Hamiltonian is not bounded below even when we consider
solutions where $q^2$ retains its sign under time evolution.
The kinetic matrix $A$ has both a positive and a negative eigenvalue,
\begin{equation}
\lambda_\pm=\frac{1}{2}\left( 1\pm\sqrt{1+4\left(\alpha q^2\right)^2}
\right)\,.
\end{equation}
Thus, we can write the Lagrangian as
\begin{equation}
L=\frac{\lambda_+}{2}(\dot{Q}^+)^2
+\frac{\lambda_-}{2}(\dot{Q}^-)^2-\tilde{V}(Q^\pm)
\end{equation}
where the variables $Q^\pm$ are defined by
\begin{equation}
\dot{Q}^\pm=u^\pm_a \dot{q}^a \,,\qquad
u^\pm=\left( \frac{\lambda_\pm}{(\alpha q^2)^2}+2 \right)^{-\frac{1}{2}}
\left(\frac{\lambda_\pm}{\alpha q^2}\,,\ 1\right)\,.
\end{equation}
Thus, the system has a positive energy and a negative energy
degree of freedom. If this model were a continuum field theory it would
be unstable, assuming the two modes interact with each other.

We expect the CRG actions corresponding to higher values of the
critical exponent $z>3$ to exhibit similar problems as the case $z=3$.
As was already noted in Sec.~\ref{sec3}, the number of time derivatives
present in the ADM representation of the action grows with $z$.
For example for $z=4,5,6$ we are most likely to recover the same
problems but in an even more complicated form than in the case $z=3$.
CRG actions corresponding to sufficiently high $z$ will necessarily
be unstable, once the order of time derivatives is greater than the
number of available constraints.

In more general perspective, we conjecture that generally covariant
higher derivative theories of gravity which aim to achieve
power-counting renormalizability via spontaneous (constraint induced)
Lorentz and/or diffeomorphism symmetry breaking will in general have to
face a similar ghost problem as CRG. The only way this could be
avoided is that either the higher time derivatives totally disappear
(cancel out) from the ADM representation of the given action or the
constraints available after symmetry breaking conspire to protect the
stability of the higher-order degrees of freedom. The former way out
would require that the spontaneous symmetry breaking reduces the
$2(d-1)$ derivatives in the invariants consisting a generally
covariant action in $d$-dimensional spacetime into $2(d-1)$ spatial
derivatives and no higher time derivatives, i.e., into an action similar
to HL gravity. This is hard to accomplish for $d=4$, as the present
work demonstrates, and likely even harder for $d>4$. The latter way out
is also hard to achieve especially when diffeomorphism invariance is
broken and hence the constraint content is no longer similar to GR.
Adding further constraints by hand is not desirable either. Still it is
not impossible that such a theory could be conceived when our
understanding on this kind of theories grows.

\section{Alternative variables for Hamiltonian formulation}\label{sec7}
Instead of \eqref{zeta_ij} alternative variables could be used to absorb
the second-order time derivatives in the action \eqref{S3.gf.ADM} in
order to construct Hamiltonian formulation of new CRG. Here we briefly
consider two more choices of higher-order variables.

\subsection{Scalar variable}
Since the second-order time derivatives in the action
\eqref{S3.gf.ADM} can be written
\begin{equation}
\begin{split}
2\left(R^{ij}+KK^{ij}\right)\partial_t \left(R_{ij}+KK_{ij}\right)&=
\partial_t
\left[\left(R^{ij}+KK^{ij}\right)\left(R_{ij}+KK_{ij}\right)\right]\\
&+4N\left(R^{ij}+KK^{ij}\right)K_{ik}g^{kl}\left(R_{lj}+KK_{lj}\right)\\
&+4\left(R^{ij}+KK^{ij}\right)D_{(i}N_{k)}g^{kl}\left(R_{lj}
+KK_{lj}\right)
\end{split}
\end{equation}
the following scalar variable can be used to absorb the second-order
time derivatives
\begin{equation}
\zeta = \left(R^{ij}+KK^{ij}\right)\left(R_{ij}+KK_{ij}\right)\,.
\end{equation}
Now the action can be replaced by
\begin{multline}\label{S3.gf.ADM.3}
S_3 = \int d^4 x \sqrt{g}N \left\{ \frac{K_{ij}K^{ij}-K^2+R}{2\kappa^2}
+ \alpha U_0 \left[ \frac{K}{N}\left( \dot{\zeta}-\cL_{\bm N}\zeta
\right) - D^iD_i \zeta \right.\right.\\
- 4\left(R^{ij}+KK^{ij}\right)K_{ik}K^{kl}
\left(R_{lj}+K K_{lj}\right) \biggr]\\
+\alpha\lambda\left[ \zeta - \left(R^{ij}+KK^{ij}\right)
\left(R_{ij}+KK_{ij}\right) \right] \biggr\} \,,
\end{multline}
where the scalar field $\lambda$ is a Lagrange multiplier.
This action has the disadvantage that it involves the extrinsic
curvature to the sixth power. As a result the momenta canonically
conjugate to the spatial metric will involve extrinsic curvature to
the fifth power. Obtaining the Hamiltonian via Legendre transformation
thus becomes a considerable problem. On the other hand, the number of
second-class constraints would be much lower than in the formalism of
Sec.~\ref{sec6}.

However, the given form of the action is suitable for illustrating the
question whether the theory contains ghosts or not.
Let us consider the following expression that contains the only time
derivative of the second-order variable $\zeta$ in the action
\begin{equation}
K\dot{\zeta}=\frac{1}{2}\begin{pmatrix}K & \dot{\zeta}\end{pmatrix}
\begin{pmatrix}
0 & 1\\
1 & 0
\end{pmatrix}
\begin{pmatrix}K \\ \dot{\zeta}\end{pmatrix}
\end{equation}
Introducing two scalar fields $\phi_1$ and $\phi_2$ as
\begin{equation}
\dot{\phi}_1= \frac{1}{\sqrt{2}}(K-\dot{\zeta})  \,, \qquad
\dot{\phi}_2=\frac{1}{\sqrt{2}}( K+\dot{\zeta}) \,,
\end{equation}
enables us to diagonalize this kinetic term. We obtain
\begin{equation}
\alpha U_0 K\dot{\zeta}=\frac{\alpha U_0}{2}
(\dot{\phi}_2^2-\dot{\phi}_1^2) \,.
 \end{equation}
Depending on the sign of the coupling constant $\alpha$ it seems that
either $\phi_1$ or $\phi_2$ has a negative kinetic term which is an
indication that given theory contains a ghost mode.
But this is just the diagonalization of the cross term between $K$ and
$\dot{\zeta}$.
We also have to substitute
\begin{equation}
K=\frac{1}{\sqrt{2}}\left( \dot{\phi}_1+\dot{\phi}_2 \right)
\,,\qquad
K_{ij}=\bar{K}_{ij}+\frac{g_{ij}}{3\sqrt{2}}\left(
\dot{\phi}_1+\dot{\phi}_2 \right)
\end{equation}
into the hole kinetic part of the action \eqref{S3.gf.ADM.3}. One
obtains that the action contains $\dot{\phi}_1$ and $\dot{\phi}_2$ up to
sixth power, including many cross terms between $\dot{\phi}_1$ and
$\dot{\phi}_2$. The sign of the kinetic terms with $\dot{\phi}_1$ and
$\dot{\phi}_2$ to powers higher than second order is $-\sgn(\alpha)$.
In particular, the terms that contain the highest powers are contained
in the term
\begin{equation}
-\frac{\alpha U_0}{54}\left( \dot{\phi}_1+\dot{\phi}_2 \right)^6 \,.
\end{equation}
Thus, when $\alpha>0$ the kinetic part does not possess a stable
minimum, while for $\alpha<0$ it does. However, answering the
question of ghosts may be more involved due to the presence of
velocities to powers higher than two. Therefore the quadratic action
\eqref{S3.gf.ADM.2} remains a safer alternative.

\subsection{Extrinsic curvature as an additional variable}
Yet another possible choice is to use the extrinsic curvature $K_{ij}$
as an additional variable
\begin{equation}
\zeta_{ij}=K_{ij}\,.
\end{equation}
Then all the time derivatives in the action \eqref{S3.gf.ADM} will be
absorbed into $\zeta_{ij}$ and $\dot{\zeta}_{ij}$. The time derivative
of the trace of the extrinsic curvature can be written
\begin{equation}
\dot{K}=g^{ij}\dot{\zeta}_{ij}-2N\zeta^{ij}\zeta_{ij}
-2\zeta^{ij}D_{(i}N_{j)}\,.
\end{equation}
The time derivative of the spatial Ricci tensor $\dot{R}_{ij}$ can also
be written in terms of $\zeta_{ij}$ by using the relation
\begin{equation}
\dot{g}_{ij}=2N\zeta_{ij}+2D_{(i}N_{j)} \,.
\end{equation}
Then the action \eqref{S3.gf.ADM} can be replaced by one that depends
on $\zeta_{ij}$, $\dot{\zeta}_{ij}$ and $g_{ij}$. Still the resulting
action is more complicated than the one based on the choice
\eqref{zeta_ij} which is used in the Hamiltonian formalism developed in
Sec.~\ref{sec6}.

\section{Conclusion}\label{sec8}
In this paper we have analyzed a new version of the so-called covariant
renormalizable gravity \cite{Kluson:2011rs} as an example of higher
derivative theory of gravity which aims to achieve power-counting
renormalizability via spontaneous (constraint induced) Lorentz and/or
diffeomorphism symmetry breaking.
It was earlier shown that this theory possesses the correct number
of degrees of freedom when fluctuations on the flat background are
analyzed and also that these fluctuations have modified dispersion
relations so that the theory is power-counting renormalizable
\cite{Kluson:2011rs}. This is a remarkable feature, because the theory
contains higher time and space derivatives which could imply potential
problems with ghosts. For that reason it would certainly be nice to
analyze the fully nonlinear theory with manifest diffeomorphism
invariance. However, due to the complexity of the action we restricted
our analysis to the case of the theory where the equation of motion
\eqref{phiconstraint} is solved for the scalar field $\phi$.
Using a foliation of spacetime defined by $\phi$ we obtained the ADM
representation of the action of CRG and its Hamiltonian structure.
It turned out that the lapse function $N$ has to depend on time only.
As a result the theory obeys the projectability condition on $N$ and
lacks a local Hamiltonian constraint, which suggests that it is not
possible to eliminate the additional scalar mode. Furthermore the fact
that the CRG action involves higher time derivatives implies that it
contains another additional degree of freedom, which can be seen when
we introduce auxiliary fields. In summary, the theory contains 4
gravitational degrees of freedom. Three of the modes are carried by the
spatial metric in conjunction with an auxiliary field $\lambda$. Another
extra scalar mode is carried by the trace component $\zeta$ of the
higher-order variable \eqref{zeta_ij} which contains the second-order
time derivatives present in the ADM representation of the action
\eqref{S3.gf.ADM}. We argued that the theory contains a degree of
freedom that carries negative energy, which will destabilize the theory
due to its interactions with positive energy degrees of freedom. We
believe the available constraints are unable to prevent this
instability, although reaching absolute certainty on this point is very
difficult because of the highly complicated form of the constraints.
The instability means that even ``empty space'' can decay into
compensating positive and negative energy excitations. Therefore this
theory cannot be considered to be a realistic description of nature,
albeit it might possess favorable renormalization characteristics. In
this respect it is similar to the generally covariant curvature-squared
gravity \cite{Stelle:1977,Stelle:1978}.

Finally we conjectured that any generally covariant higher derivative
theory of gravity which aims to achieve power-counting
renormalizability via spontaneous (constraint induced)
Lorentz and/or diffeomorphism symmetry breaking will have to face a
similar challenge with ghosts as CRG. This is because of the
difficulties in reducing the number of time derivatives in a generally
covariant higher-order theory via spontaneous  symmetry breaking and the
reduced structure of constraints after the symmetry has been broken.
It remains to be seen whether this challenge can be overcome.

We should emphasize that the secondary constraints of the theory are
very complicated. Especially the second-class secondary constraints
turned out inconveniently complicated. As a result the Hamiltonian has a
very complicated form. Since the canonical variables have such
extremely involved relations, the description of general dynamics is
very complicated. Canonical quantization of the theory would certainly
be a very hard task. Path integral quantization might be a more feasible
approach.

Very recently we have begun to suspect that in this type of theory it
might be better to treat the normal $n^\mu$ as a genuine dynamical
variable with constraints restricting it to unit norm and zero
vorticity. This way a more a general treatment could perhaps be
achieved. In covariant form the action would be defined as
\begin{multline}
S_3 = \int d^4 x \sqrt{-\gM} \left[ \frac{\RM}{2\kappa^2} - \alpha
\left( R^{\alpha\beta} + KK^{\alpha\beta} - D^\alpha a^\beta \right)
\right.\\
\times \left( n^\mu n^\nu \nabla_\mu \nabla_\nu + \nabla^\mu \nabla_\mu
\right) \left( R_{\alpha\beta} + KK_{\alpha\beta} - D_\alpha a_\beta
\right)\\
+ \lambda\left( n_\mu n^\mu+1 \right)
+ B^{\mu\nu}\cF_{\mu\nu}
+ M_{\mu\nu\rho\sigma}B^{\mu\nu}B^{\rho\sigma} \Bigr],
\end{multline}
where the vorticity for $n_\mu$ is
\begin{equation}
\cF_{\mu\nu}=\projector{\rho}{\mu}\projector{\sigma}{\nu}\nabla_{[\rho}
n_{\sigma]}
\end{equation}
and $\lambda$, $B^{\mu\nu}$ and $M_{\mu\nu\rho\sigma}$ are Lagrange
multiplier fields. Variations of the action with respect to the Lagrange
multipliers yield
\begin{equation}
n_\mu n^\mu=-1,\quad
B^{\mu\nu}=0,\quad
\cF_{\mu\nu}=0.
\end{equation}
The normal would be associated with the scalar field $\phi$ by
\begin{equation}
n_\mu=-N\nabla_\mu \phi
\end{equation}
and choosing it to be the time $\phi=t$ (ADM gauge choice) yields the
ADM formulation of the theory.
This approach would be similar to the covariant formulation of HL
gravity \cite{Germani:2009} (also see \cite{Blas:2009,Blas:2010a}), but
with a much more complicated action: higher-order time derivatives and
several extra kinetic terms etc. Indeed, the only conceivable advantage
of CRG over HL gravity is the spontaneous breaking of Lorentz
invariance in the high energy regime. The idea of achieving
renormalizable gravity via spontaneous symmetry breaking is
certainly appealing, but we doubt whether it is worth the price of
accepting such a complicated Hamiltonian structure, let alone an
unstable extra degree of freedom which is simply unacceptable.

\paragraph{Acknowledgements}
The support of the Academy of Finland under Projects No. 136539
and No. 140886 is gratefully acknowledged.
The work of J.K. was supported by the Grant agency of the Czech
republic under Grant No. P201/12/G028.
M.O. is supported by the Jenny and Antti Wihuri Foundation.

\appendix
\section{Appendix}

\subsection{Symmetrization}\label{appendix1}
Symmetrization and antisymmetrization of tensor indices is denoted by
parentheses and brackets, respectively, e.g.,
\begin{equation*}
B_{(\alpha\beta)} = \frac{1}{2}\left( B_{\alpha\beta}+B_{\beta\alpha}
\right) \,,\qquad B_{[\alpha\beta]} = \frac{1}{2}\left(
B_{\alpha\beta}-B_{\beta\alpha} \right)\,.
\end{equation*}
We may also use the following notation if it is more convenient
\begin{equation*}
B_{\alpha\beta}+(\alpha\leftrightarrow\beta) = 2B_{(\alpha\beta)}
\,,\qquad B_{\alpha\beta}-(\alpha\leftrightarrow\beta) =
2B_{[\alpha\beta]} \,,
\end{equation*}
as can be the case in long expressions with many terms.

\subsection{Decomposition of the Riemann tensor}\label{appendix2}
Decomposition of the Riemann tensor of spacetime into components tangent
and normal to the spatial hypersurfaces $\Sigma_t$ is given by the
following identities:
\begin{enumerate}
\item Gauss relation
\begin{equation}\label{GaussEq}
\projector{\gamma}{\mu}\projector{\nu}{\delta}\projector{\rho}{\alpha}
\projector{\sigma}{\beta} \RM^{\mu}_{\phantom\mu\nu\rho\sigma} =
R^{\gamma}_{\phantom\gamma\delta\alpha\beta} +
K^{\gamma}_{\phantom\gamma\alpha}K_{\delta\beta} -
K^{\gamma}_{\phantom\gamma\beta}K_{\alpha\delta} \,.
\end{equation}

\item Codazzi relation
\begin{equation}\label{CodazziEq}
\projector{\gamma}{\mu}n^\nu
\projector{\rho}{\alpha}\projector{\sigma}{\beta}
\RM^{\mu}_{\phantom\mu\nu\rho\sigma} =
2D_{[\alpha}K^{\gamma}_{\phantom\gamma\beta]} \,.
\end{equation}

\item Ricci equation
\begin{equation}\label{RicciEq}
g_{\alpha\mu}n^\nu \projector{\rho}{\beta}n^\sigma
\RM^{\mu}_{\phantom\mu\nu\rho\sigma} =
K_{\alpha\mu}K^{\mu}_{\phantom\mu\beta} + \frac{1}{N}D_\alpha D_\beta N
- \frac{1}{N}\cL_{Nn} K_{\alpha\beta} \,.
\end{equation}
\end{enumerate}

These imply for example the decomposition of the scalar curvature of
spacetime
\begin{equation}\label{RM}
\RM = R + K_{ij}K^{ij} - K^2 + 2\nabla_\mu\left(n^\mu K\right) -
\frac{2}{N}D^i D_i N \,,
\end{equation}
where
\[
2\nabla_\mu\left(n^\mu K\right) = 2K^2 + \frac{2}{N}\left(
\dot{K}-N^i\p_i K \right)\,.
\]

\subsection{Decomposition of covariant derivatives of tensors tangent
to spatial hypersurfaces}\label{appendix3}

Here we shall obtain spacetime decompositions for first- and
second-order covariant derivatives of tensors tangent to spatial
hypersurfaces $\Sigma_t$.

In the following calculations we will frequently need the covariant
derivative of the orthogonal projector onto the spatial hypersurface
\begin{equation}
\nabla_\mu \projector{\nu}{\alpha} = \left( K_\mu^{\phantom\mu\nu} -
n_\mu a^\nu \right) n_\alpha + n^\nu \left( K_{\mu\alpha} - n_\mu
a_\alpha \right) \,.
\end{equation}

Consider a symmetric rank 2 covariant tensor field $A_{\alpha\beta}$
that is tangent to $\Sigma_t$, such as the two tensor fields inside the
parentheses in \eqref{covdertodecompose}. Such a tensor field is
invariant under  the orthogonal projector,
$A_{\alpha\beta}=\projector{\mu}{\alpha}\projector{\nu}{\beta}
A_{\mu\nu}$,
since $n^{\alpha}A_{\alpha\beta}=n^{\alpha}A_{\beta\alpha}=0$.
Let us decompose the covariant derivative of
$A_{\alpha\beta}$:
\begin{equation}\label{decompNablaA}
\begin{split}
\nabla_\mu A_{\alpha\beta} &= \delta^\nu_\mu \nabla_\nu \left(
\projector{\rho}{\alpha}\projector{\sigma}{\beta} A_{\rho\sigma} \right)
= \left( \projector{\nu}{\mu} - n^\nu n_\mu \right)
\projector{\rho}{\alpha}\projector{\sigma}{\beta} \nabla_\nu
A_{\rho\sigma} + \nabla_\mu \left(
\projector{\rho}{\alpha}\projector{\sigma}{\beta} \right) A_{\rho\sigma}
\\
&=
\projector{\nu}{\mu}\projector{\rho}{\alpha}\projector{\sigma}{\beta}
\nabla_\nu  A_{\rho\sigma} - n_\mu
\projector{\rho}{\alpha}\projector{\sigma}{\beta}\nabla_n
A_{\rho\sigma} + \left( \nabla_\mu
\projector{\rho}{\alpha}\projector{\sigma}{\beta} +
\projector{\rho}{\alpha} \nabla_\mu \projector{\sigma}{\beta} \right)
A_{\rho\sigma} \\
&= D_\mu A_{\alpha\beta} - n_\mu \left(
\frac{1}{N}\cL_{Nn}A_{\alpha\beta} -
2K_{(\alpha}^{\phantom{(\alpha}\nu}A_{\beta)\nu} \right) + 2\left(
K_\mu^{\phantom{\mu}\nu} - n_\mu a^\nu \right) n_{(\alpha}A_{\beta)\nu}
\\
&= D_\mu A_{\alpha\beta} - n_\mu \left(
\frac{1}{N}\cL_{Nn}A_{\alpha\beta} -
2K_{(\alpha}^{\phantom{(\alpha}\nu}A_{\beta)\nu} + 2
n_{(\alpha}A_{\beta)\nu}a^\nu \right) + 2K_\mu^{\phantom{\mu}\nu}
n_{(\alpha}A_{\beta)\nu}
\end{split}
\end{equation}
The component of $\nabla_\mu A_{\alpha\beta}$ that is fully tangent to
$\Sigma_t$ is simply $
\projector{\nu}{\mu}\projector{\rho}{\alpha}\projector{\sigma}{\beta}
\nabla_\nu  A_{\rho\sigma}=D_\mu A_{\alpha\beta}$. For completeness let
us list the remaining seven components of the decomposition of
$\nabla_\mu A_{\alpha\beta}$:
\begin{align*}
-n_\mu && n^\nu
\projector{\rho}{\alpha}\projector{\sigma}{\beta}\nabla_\nu
A_{\rho\sigma} &= \frac{1}{N}\cL_{Nn}A_{\alpha\beta} -
2K_{(\alpha}^{\phantom{(\alpha}\nu}A_{\beta)\nu} \,,\\
-n_\alpha &&
\projector{\nu}{\mu}n^{\rho}\projector{\sigma}{\beta}\nabla_\nu
A_{\rho\sigma} &= - K_\mu^{\phantom{\mu}\nu} A_{\nu\beta} \,,\\
-n_\beta &&
\projector{\nu}{\mu}\projector{\rho}{\alpha}n^{\sigma}\nabla_\nu
A_{\rho\sigma} &= - K_\mu^{\phantom{\mu}\nu} A_{\alpha\nu} \,,\\
+n_\mu n_\alpha && n^\nu n^{\rho}\projector{\sigma}{\beta}\nabla_\nu
A_{\rho\sigma} &= - a^{\nu}A_{\nu\beta} \,,\\
+n_\mu n_\beta && n^\nu \projector{\rho}{\alpha}n^{\sigma}\nabla_\nu
A_{\rho\sigma} &= - a^{\nu}A_{\alpha\nu} \,,\\
+n_\alpha n_\beta && \projector{\nu}{\mu}n^{\rho}n^{\sigma}\nabla_\nu
A_{\rho\sigma} &= 0 \,,\\
-n_\mu n_\alpha n_\beta && n^\nu n^{\rho}n^{\sigma}\nabla_\nu
A_{\rho\sigma} &= 0 \,,
\end{align*}
where in the left-hand side column we present the part of each component
that is normal to $\Sigma_t$ in the decomposition \eqref{decompNablaA}.

For future needs we generalize Eqs. \eqref{LieNnK}, \eqref{LieNnK2} and
\eqref{decompNablaA} for a generic covariant tensor field
$B_{\alpha_1\cdots\alpha_k}$ that is tangent to $\Sigma_t$:
\begin{equation}\label{LieNnB}
\frac{1}{N}\cL_{Nn}B_{\alpha_1\cdots\alpha_k} = \nabla_n
B_{\alpha_1\cdots\alpha_k} + \left( K_{\alpha_1}^{\phantom{\alpha_1}\mu}
- n_{\alpha_1} a^\mu \right)  B_{\mu\cdots\alpha_k} + \cdots + \left(
K_{\alpha_k}^{\phantom{\alpha_k}\mu} - n_{\alpha_k} a^\mu \right)
B_{\alpha_1\cdots\mu} \,,
\end{equation}
\begin{equation}\label{LieNnB2}
\frac{1}{N}\cL_{Nn}B_{\alpha_1\cdots\alpha_k} =
\projector{\mu_1}{\alpha_1}\cdots\projector{\mu_k}{\alpha_k} \nabla_n
B_{\mu_1\cdots\mu_k} +
K_{\alpha_1}^{\phantom{\alpha_1}\mu}B_{\mu\cdots\alpha_k} + \cdots +
K_{\alpha_k}^{\phantom{\alpha_k}\mu}B_{\alpha_1\cdots\mu} \,,
\end{equation}
\begin{equation}\label{decompNablaB}
\begin{split}
\nabla_\mu B_{\alpha_1\cdots\alpha_k} &= D_\mu
B_{\alpha_1\cdots\alpha_k} - n_\mu \left(
\frac{1}{N}\cL_{Nn}B_{\alpha_1\cdots\alpha_k} -
K_{\alpha_1}^{\phantom{\alpha_1}\nu}B_{\nu\cdots\alpha_k} - \cdots -
K_{\alpha_k}^{\phantom{\alpha_k}\nu}B_{\alpha_1\cdots\nu} \right. \\
&+ n_{\alpha_1}a^\nu B_{\nu\cdots\alpha_k} + \cdots + n_{\alpha_k}a^\nu
B_{\alpha_1\cdots\nu} \Bigr)  + n_{\alpha_1} K_\mu^{\phantom{\mu}\nu}
B_{\nu\cdots\alpha_k} + \cdots + n_{\alpha_k} K_\mu^{\phantom{\mu}\nu}
B_{\alpha_1\cdots\nu} \,.
\end{split}
\end{equation}
It is worth noting that, due to the implication $a^{\mu}=0$ of the
condition $N=N(t)$ [see \eqref{a=0}], in \eqref{LieNnB} $\nabla_n
B_{\alpha_1\cdots\alpha_k}$ is already tangent to $\Sigma_t$, and hence
the projection of \eqref{LieNnB} in \eqref{LieNnB2} is trivial. However,
as we did in \eqref{LieNnK} and \eqref{decompNablaA}, we continue to
write $a^{\mu}$ explicitly in the decompositions of the covariant
derivatives, in order to retain generality in this point.

Then consider the second-order covariant derivative of
$A_{\alpha\beta}$. We decompose the second-order covariant derivative of
$A_{\alpha\beta}$ similarly as the first-order covariant derivative in
\eqref{decompNablaA}:
\begin{equation}\label{decompNabla2A}
\begin{split}
\nabla_\mu \nabla_\nu A_{\alpha\beta} &= \delta^\rho_\mu
\delta^\sigma_\nu \nabla_\rho \nabla_\sigma \left(
\projector{\lambda}{\alpha}\projector{\tau}{\beta} A_{\lambda\tau}
\right) \\\
&= \left( \projector{\rho}{\mu} - n^\rho n_\mu \right) \left(
\projector{\sigma}{\nu} - n^\sigma n_\nu \right)
\projector{\lambda}{\alpha}\projector{\tau}{\beta} \nabla_\rho
\nabla_\sigma A_{\lambda\tau} \\
&+ 2\nabla_{(\mu|} \left(
\projector{\lambda}{\alpha}\projector{\tau}{\beta} \right)
\nabla_{|\nu)} A_{\lambda\tau} + \nabla_\mu \nabla_\nu \left(
\projector{\lambda}{\alpha}\projector{\tau}{\beta} \right)
A_{\lambda\tau} \\
&= \projector{\rho}{\mu}
\projector{\sigma}{\nu}\projector{\lambda}{\alpha}\projector{\tau}{\beta
}\nabla_\rho \nabla_\sigma  A_{\lambda\tau} - n_\mu n^\rho
\projector{\sigma}{\nu}
\projector{\lambda}{\alpha}\projector{\tau}{\beta} \nabla_\rho
\nabla_\sigma A_{\lambda\tau} \\
&- n_\nu \projector{\rho}{\mu} n^\sigma
\projector{\lambda}{\alpha}\projector{\tau}{\beta} \nabla_\rho
\nabla_\sigma A_{\lambda\tau} + n_\mu n_\nu n^\rho n^\sigma
\projector{\lambda}{\alpha}\projector{\tau}{\beta} \nabla_\rho
\nabla_\sigma A_{\lambda\tau} \\
&+ 2\nabla_{(\mu|} \left(
\projector{\rho}{\alpha}\projector{\sigma}{\beta} \right) \nabla_{|\nu)}
A_{\rho\sigma} + \nabla_\mu \nabla_\nu \left(
\projector{\rho}{\alpha}\projector{\sigma}{\beta} \right) A_{\rho\sigma}
\,.
\end{split}
\end{equation}
However, now we still have to decompose all the terms in
\eqref{decompNabla2A}.
In order to obtain the component of $\nabla_\mu\nabla_\nu
A_{\alpha\beta}$ that is fully tangent to $\Sigma_t$ we use \eqref{DmuT}
to calculate
\begin{equation*}
\begin{split}
D_\mu D_\nu A_{\alpha\beta} &=
\projector{\rho}{\mu}\projector{\sigma}{\nu}\projector{\lambda}{\alpha}
\projector{\tau}{\beta} \nabla_\rho D_\sigma A_{\lambda\tau} =
\projector{\rho}{\mu}\projector{\sigma}{\nu}\projector{\lambda}{\alpha}
\projector{\tau}{\beta} \nabla_\rho \left( \projector{\phi}{\sigma}
\projector{\varphi}{\lambda} \projector{\chi}{\tau} \nabla_\phi
A_{\varphi\chi} \right) \\
&=
\projector{\rho}{\mu}\projector{\sigma}{\nu}\projector{\lambda}{\alpha}
\projector{\tau}{\beta} \nabla_\rho \left( \projector{\phi}{\sigma}
\projector{\varphi}{\lambda} \projector{\chi}{\tau} \right) \nabla_\phi
A_{\varphi\chi} +
\projector{\rho}{\mu}\projector{\sigma}{\nu}\projector{\lambda}{\alpha}
\projector{\tau}{\beta} \nabla_\rho \nabla_\sigma A_{\lambda\tau} \,,
\end{split}
\end{equation*}
which together with \eqref{decompNablaA} finally gives
\begin{equation}\label{perpNabla2A}
\projector{\rho}{\mu}\projector{\sigma}{\nu}\projector{\lambda}{\alpha}
\projector{\tau}{\beta} \nabla_\rho \nabla_\sigma A_{\lambda\tau} =
D_\mu D_\nu A_{\alpha\beta} - K_{\mu\nu}\left(
\frac{1}{N}\cL_{Nn}A_{\alpha\beta} -
2K_{(\alpha}^{\phantom{(\alpha}\rho}A_{\beta)\rho} \right) +
2K_{\mu(\alpha|}K_\nu^{\phantom{\nu}\rho}A_{\rho|\beta)} \,.
\end{equation}
Next consider the $-n_\mu$ component of the decomposition of $\nabla_\mu
\nabla_\nu A_{\alpha\beta}$ [the second term in \eqref{decompNabla2A}].
It can be decomposed by using \eqref{DmuT} in order to write
\begin{equation}\label{Nabla_nDA}
\nabla_n D_\nu A_{\alpha\beta} = \nabla_n \left(
\projector{\sigma}{\nu}\projector{\lambda}{\alpha}\projector{\tau}{\beta
} \nabla_\sigma A_{\lambda\tau} \right) = \nabla_n \left(
\projector{\sigma}{\nu}\projector{\lambda}{\alpha}\projector{\tau}{\beta
} \right) \nabla_\sigma A_{\lambda\tau} + n^\rho
\projector{\sigma}{\nu}\projector{\lambda}{\alpha}\projector{\tau}{\beta
} \nabla_\rho \nabla_\sigma A_{\lambda\tau} \,,
\end{equation}
where the last term is what we are looking to decompose. Decomposition
of the left-hand side of \eqref{Nabla_nDA} is given by \eqref{LieNnB}
\begin{equation*}
\begin{split}
\nabla_n D_\nu A_{\alpha\beta} &= \frac{1}{N}\cL_{Nn} D_\nu
A_{\alpha\beta} - \left( K_{\nu}^{\phantom{\nu}\rho} - n_{\nu} a^\rho
\right) D_\rho A_{\alpha\beta} \\
&- \left( K_{\alpha}^{\phantom{\alpha}\rho} - n_{\alpha} a^\rho \right)
D_\nu A_{\rho\beta} - \left( K_{\beta}^{\phantom{\beta}\rho} - n_{\beta}
a^\rho \right) D_\nu A_{\alpha\rho} \,.
\end{split}
\end{equation*}
Decomposition of the first term in the right-hand side of
\eqref{Nabla_nDA} is given by \eqref{decompNablaA} and taking the
derivative of the projectors,
\begin{equation*}
\begin{split}
\nabla_n \left(
\projector{\sigma}{\nu}\projector{\lambda}{\alpha}\projector{\tau}{\beta
} \right) \nabla_\sigma A_{\lambda\tau} &= n_\nu a^\rho D_\rho
A_{\alpha\beta}
+a_\nu \left( \frac{1}{N}\cL_{Nn}A_{\alpha\beta} -
2K_{(\alpha}^{\phantom{(\alpha}\rho}A_{\beta)\rho} \right)\\
&+ 2n_{(\alpha|}D_\nu A_{|\beta)\rho}a^\rho -
2K_\nu^{\phantom\nu\rho}a_{(\alpha}A_{\beta)\rho} \,.
\end{split}
\end{equation*}
Thus, we obtain the $-n_\mu$ component of the decomposition
\eqref{decompNabla2A} of $\nabla_\mu \nabla_\nu A_{\alpha\beta}$:
\begin{equation}\label{comp_n_mu}
\begin{split}
n^\rho \projector{\sigma}{\nu} \projector{\lambda}{\alpha}
\projector{\tau}{\beta} \nabla_\rho \nabla_\sigma A_{\lambda\tau} &=
\frac{1}{N}\cL_{Nn} D_\nu A_{\alpha\beta} - K_{\nu}^{\phantom{\nu}\rho}
D_\rho A_{\alpha\beta} - 2K_{(\alpha|}^{\phantom{(\alpha|}\rho} D_\nu
A_{|\beta)\rho} \\
&- a_\nu \left( \frac{1}{N}\cL_{Nn}A_{\alpha\beta} -
2K_{(\alpha}^{\phantom{(\alpha}\rho}A_{\beta)\rho} \right) +
2K_\nu^{\phantom\nu\rho}a_{(\alpha}A_{\beta)\rho} \,.
\end{split}
\end{equation}
The $-n_\nu$ component of the decomposition of $\nabla_\mu \nabla_\nu
A_{\alpha\beta}$ [the third term in \eqref{decompNabla2A}] can be
written
\begin{equation}
\projector{\rho}{\mu} n^\sigma
\projector{\lambda}{\alpha}\projector{\tau}{\beta} \nabla_\rho
\nabla_\sigma A_{\lambda\tau} =  n^\rho \projector{\sigma}{\mu}
\projector{\lambda}{\alpha} \projector{\tau}{\beta} \nabla_\rho
\nabla_\sigma A_{\lambda\tau} + \projector{\rho}{\mu} n^\sigma
\projector{\lambda}{\alpha}\projector{\tau}{\beta} [\nabla_\rho,
\nabla_\sigma] A_{\lambda\tau} \,.
\end{equation}
In the second term we use the \emph{Ricci identity} and the fact that
$A_{\alpha\beta}$ is tangent to $\Sigma_t$ to obtain
\begin{equation*}
\begin{split}
\projector{\rho}{\mu} n^\sigma
\projector{\lambda}{\alpha}\projector{\tau}{\beta} [\nabla_\rho,
\nabla_\sigma] A_{\lambda\tau}
&= - \projector{\rho}{\mu} n^\sigma
\projector{\lambda}{\alpha}\projector{\tau}{\beta} \left(
\RM^\phi_{\phantom\phi\lambda\rho\sigma} A_{\phi\tau} +
\RM^\phi_{\phantom\phi\tau\rho\sigma} A_{\lambda\phi} \right)\\
&= - \projector{\rho}{\mu} n^\sigma
\projector{\lambda}{\alpha}\projector{\tau}{\beta} g^{\phi\varphi}
\left( \RM_{\varphi\lambda\rho\sigma} A_{\phi\tau} +
\RM_{\varphi\tau\rho\sigma} A_{\lambda\phi} \right)\\
&= - g_{\mu\rho} n^\sigma \projector{\varphi}{\phi} \left(
\projector{\lambda}{\alpha}
\RM^{\rho}_{\phantom\rho\sigma\varphi\lambda}A^\phi_{\phantom\phi\beta}
+ \projector{\tau}{\beta}
\RM^{\rho}_{\phantom\rho\sigma\varphi\tau}A_\alpha^{\phantom\alpha\phi}
\right)\\
&= - 2D_{[\rho|}K_{\mu|\alpha]}A^\rho_{\phantom\rho\beta} -
2D_{[\rho|}K_{\mu|\beta]}A_\alpha^{\phantom\alpha\rho} \\
&= 2D_{(\alpha|}K_{\mu\rho}A^\rho_{\phantom\rho|\beta)} -
2D_{\rho}K_{\mu(\alpha}A^\rho_{\phantom\rho\beta)}
\end{split}
\end{equation*}
where we have used the Codazzi relation \eqref{CodazziEq} in the second
last equality. Thus, the $-n_\nu$ component of the decomposition
\eqref{decompNabla2A} of $\nabla_\mu \nabla_\nu A_{\alpha\beta}$ is
\begin{equation}\label{comp_n_nu}
\begin{split}
\projector{\rho}{\mu} n^\sigma
\projector{\lambda}{\alpha}\projector{\tau}{\beta} \nabla_\rho
\nabla_\sigma A_{\lambda\tau}
&= \frac{1}{N}\cL_{Nn} D_\mu A_{\alpha\beta} -
K_{\mu}^{\phantom{\mu}\rho} D_\rho A_{\alpha\beta} -
2K_{(\alpha|}^{\phantom{(\alpha|}\rho} D_\mu A_{|\beta)\rho} \\
&- a_\mu \left( \frac{1}{N}\cL_{Nn}A_{\alpha\beta} -
2K_{(\alpha}^{\phantom{(\alpha}\rho}A_{\beta)\rho} \right) +
2K_\mu^{\phantom\mu\rho}a_{(\alpha}A_{\beta)\rho}\\
&+ 2D_{(\alpha|}K_{\mu\rho}A^\rho_{\phantom\rho|\beta)} -
2D_{\rho}K_{\mu(\alpha}A^\rho_{\phantom\rho\beta)} \,.
\end{split}
\end{equation}
Next consider the $+n_\mu n_\nu$ component of the decomposition of
$\nabla_\mu \nabla_\nu A_{\alpha\beta}$ [the fourth term in
\eqref{decompNabla2A}]. Let us write
\begin{equation}
\nabla_n \left( \projector{\lambda}{\alpha}\projector{\tau}{\beta}
\nabla_n A_{\lambda\tau} \right) = \nabla_n \left( n^\sigma
\projector{\lambda}{\alpha}\projector{\tau}{\beta} \right) \nabla_\sigma
A_{\lambda\tau} + n^\rho n^\sigma
\projector{\lambda}{\alpha}\projector{\tau}{\beta} \nabla_\rho
\nabla_\sigma A_{\lambda\tau} \,,
\end{equation}
where the last term is what we are looking to decompose. The left-hand
side can alternatively be written by using \eqref{LieNnB} and
\eqref{LieNnB2}:
\begin{equation*}
\begin{split}
\nabla_n \left( \projector{\lambda}{\alpha}\projector{\tau}{\beta}
\nabla_n A_{\lambda\tau} \right) &= \nabla_n \left( \frac{1}{N}\cL_{Nn}
A_{\alpha\beta} - 2K_{(\alpha}^{\phantom{(\alpha}\rho}A_{\beta)\rho}
\right) \\
&= \frac{1}{N}\cL_{Nn} \left( \frac{1}{N}\cL_{Nn} A_{\alpha\beta} -
2K_{(\alpha}^{\phantom{(\alpha}\rho}A_{\beta)\rho} \right) \\
&- \left( K_{\alpha}^{\phantom{\alpha}\rho} - n_\alpha a^\rho \right)
\left( \frac{1}{N}\cL_{Nn} A_{\rho\beta} -
2K_{(\rho}^{\phantom{(\rho}\sigma}A_{\beta)\sigma} \right) \\
&- \left( K_{\beta}^{\phantom{\beta}\rho} - n_\beta a^\rho \right)
\left( \frac{1}{N}\cL_{Nn} A_{\alpha\rho} -
K_{(\alpha}^{\phantom{(\alpha}\sigma}A_{\rho)\sigma} \right) \,.
\end{split}
\end{equation*}
We also need to compute
\begin{equation*}
\begin{split}
\nabla_n \left( n^\sigma
\projector{\lambda}{\alpha}\projector{\tau}{\beta} \right) \nabla_\sigma
A_{\lambda\tau} &= a^\rho D_\rho A_{\alpha\beta} + n_\alpha a^\rho
\left( \frac{1}{N}\cL_{Nn} A_{\rho\beta} -
2K_{(\rho}^{\phantom{(\rho}\sigma}A_{\beta)\sigma} \right)\\
&+n_\beta a^\rho \left( \frac{1}{N}\cL_{Nn} A_{\alpha\rho} -
K_{(\alpha}^{\phantom{(\alpha}\sigma}A_{\rho)\sigma} \right)
- 2a_{(\alpha}A_{\beta)\rho}a^{\rho} \,.
\end{split}
\end{equation*}
Thus, we obtain the $+n_\mu n_\nu$ component of the decomposition
\eqref{decompNabla2A} of $\nabla_\mu \nabla_\nu A_{\alpha\beta}$:
\begin{equation}
\begin{split}
n^\rho n^\sigma \projector{\lambda}{\alpha}\projector{\tau}{\beta}
\nabla_\rho \nabla_\sigma A_{\lambda\tau}
&= \frac{1}{N}\cL_{Nn} \left( \frac{1}{N}\cL_{Nn} A_{\alpha\beta}
\right) - \frac{2}{N}\cL_{Nn}
K_{(\alpha}^{\phantom{(\alpha}\rho}A_{\beta)\rho} \\
&- 4K_{(\alpha|}^{\phantom{(\alpha|}\rho} \frac{1}{N}\cL_{Nn}
A_{|\beta)\rho}
+ 2K_{(\alpha}^{\phantom{(\alpha}\rho}
K_{\beta)}^{\phantom{\beta)}\sigma} A_{\rho\sigma} +
2K_{(\alpha|}^{\phantom{(\alpha|}\rho} K_{\rho}^{\phantom{\rho}\sigma}
A_{|\beta)\sigma}\\
&- a^\rho D_\rho A_{\alpha\beta} + 2a_{(\alpha}A_{\beta)\rho}a^{\rho}
\,.
\end{split}
\end{equation}
The remaining 12 components of the decomposition of $\nabla_\mu
\nabla_\nu A_{\alpha\beta}$ are given in the last two terms of Eq.
\eqref{decompNabla2A}. Let us calculate them:
\begin{equation}\label{restofdecomp1}
\begin{split}
2\nabla_{(\mu|} \left( \projector{\rho}{\alpha}\projector{\sigma}{\beta}
\right) \nabla_{|\nu)} A_{\rho\sigma} &= n_\alpha \left[
K_\mu^{\phantom\mu\rho} D_\nu A_{\rho\beta} - n_\mu \left( a^\rho D_\nu
A_{\rho\beta} + K_\nu^{\phantom\nu\rho} \frac{1}{N}\cL_{Nn}A_{\rho\beta}
\right.\right. \\
&-\left.
K_\nu^{\phantom\nu\rho}K_\rho^{\phantom\rho\sigma}A_{\sigma\beta} -
K_\nu^{\phantom\nu\rho}K_\beta^{\phantom\beta\sigma}A_{\rho\sigma}
\right) + n_\mu n_\nu  \left( a^{\rho}\frac{1}{N}\cL_{Nn}A_{\rho\beta}
\right.\\
&-\left. a^{\rho}K_\rho^{\phantom\rho\sigma}A_{\sigma\beta} -
a^{\rho}K_\beta^{\phantom\beta\sigma}A_{\rho\sigma} \right) +
(\mu\leftrightarrow\nu) \biggr]  \\
&+ \left[ \left( K_\mu^{\phantom\mu\rho} - n_\mu a^\rho \right) \left(
K_{\nu\alpha} - n_\nu a_\alpha \right) + (\mu\leftrightarrow\nu) \right]
A_{\rho\beta} + (\alpha\leftrightarrow\beta) \,,
\end{split}
\end{equation}
\begin{equation}\label{restofdecomp2}
\begin{split}
\nabla_\mu \nabla_\nu \left(
\projector{\rho}{\alpha}\projector{\sigma}{\beta} \right) A_{\rho\sigma}
&= \nabla_\mu \nabla_\nu \projector{\rho}{\alpha} A_{\rho\beta} +
\nabla_{\mu} \projector{\rho}{\alpha} \nabla_{\nu}
\projector{\sigma}{\beta} A_{\rho\sigma} +
(\alpha\leftrightarrow\beta)\\
&= n_\alpha \left[ D_\mu K_{\nu\rho} - K_{\mu\nu}a_\rho - n_\mu \left(
\frac{1}{N}\cL_{Nn} K_{\nu\rho} -
2K_{\nu\sigma}K^{\sigma}_{\phantom\sigma\rho} - a_\nu a_\rho \right)
\right.\\
&-\left. n_\nu \left( D_\mu a_\rho -
K_\mu^{\phantom\mu\sigma}K_{\sigma\rho} \right)
+ n_\mu n_\nu \left( \frac{1}{N}\cL_{Nn} a_\rho - 2
K_{\rho\sigma}a^\sigma \right) \right] A^{\rho}_{\phantom\rho\beta}\\
&+ \left[ \left( K_\mu^{\phantom\mu\rho} - n_\mu a^\rho \right) \left(
K_{\nu\alpha} - n_\nu a_\alpha \right) + (\mu\leftrightarrow\nu) \right]
A_{\rho\beta} \\
&+ n_\alpha n_\beta \left( K_\mu^{\phantom\mu\rho} - n_\mu a^\rho
\right) \left( K_\nu^{\phantom\nu\sigma} - n_\nu a^\sigma \right)
A_{\rho\sigma} + (\alpha\leftrightarrow\beta) \,.
\end{split}
\end{equation}
Taking the sum of \eqref{restofdecomp1} and \eqref{restofdecomp2} gives
\begin{equation}\label{restofdecomp}
\begin{split}
2\nabla_{(\mu|} \left( \projector{\rho}{\alpha}\projector{\sigma}{\beta}
\right) & \nabla_{|\nu)} A_{\rho\sigma} + \nabla_\mu \nabla_\nu \left(
\projector{\rho}{\alpha}\projector{\sigma}{\beta} \right) A_{\rho\sigma}
\\
&=  n_\alpha \Bigl[ K_\mu^{\phantom\mu\rho} D_\nu A_{\rho\beta} +
K_\nu^{\phantom\nu\rho} D_\mu A_{\rho\beta} + D_\mu
K_{\nu\rho}A^{\rho}_{\phantom\rho\beta} - K_{\mu\nu}a^\rho A_{\rho\beta}
 \\
&- n_\mu \left( a^\rho D_\nu A_{\rho\beta} + K_\nu^{\phantom\nu\rho}
\frac{1}{N}\cL_{Nn}A_{\rho\beta} + \frac{1}{N}\cL_{Nn}
K_{\nu\rho}A^{\rho}_{\phantom\rho\beta} \right.\\
&-\left.
3K_\nu^{\phantom\nu\rho}K_\rho^{\phantom\rho\sigma}A_{\sigma\beta} -
K_\nu^{\phantom\nu\rho}K_\beta^{\phantom\beta\sigma}A_{\rho\sigma} -
a_\nu a^\rho A_{\rho\beta} \right) \\
&- n_\nu \left( a^\rho D_\mu A_{\rho\beta} + D_\mu a_\rho
A^{\rho}_{\phantom\rho\beta} + K_\mu^{\phantom\mu\rho}
\frac{1}{N}\cL_{Nn}A_{\rho\beta} \right.\\
&-\left.
2K_\mu^{\phantom\mu\rho}K_\rho^{\phantom\rho\sigma}A_{\sigma\beta} -
K_\mu^{\phantom\mu\rho}K_\beta^{\phantom\beta\sigma}A_{\rho\sigma}
\right) \\
&+ n_\mu n_\nu 2 \left( a^{\rho}\frac{1}{N}\cL_{Nn}A_{\rho\beta} +
\frac{1}{2N}\cL_{Nn} a_\rho A^{\rho}_{\phantom\rho\beta} -
2a^{\rho}K_\rho^{\phantom\rho\sigma}A_{\sigma\beta} \right.\\
&-\left. a^{\rho}K_\beta^{\phantom\beta\sigma}A_{\rho\sigma} \right)
\Bigr]
+ n_\alpha n_\beta \left( K_\mu^{\phantom\mu\rho} - n_\mu a^\rho \right)
\left( K_\nu^{\phantom\nu\sigma} - n_\nu a^\sigma \right) A_{\rho\sigma}
\\
&+ (\alpha\leftrightarrow\beta) \,.
\end{split}
\end{equation}
Now we can read the rest of the components of the decomposition of
$\nabla_\mu \nabla_\nu A_{\alpha\beta}$ from \eqref{restofdecomp}:
\begin{align*}
-n_\alpha &&
\projector{\rho}{\mu}\projector{\sigma}{\nu}n^{\lambda}\projector{\tau}{
\beta}\nabla_\rho \nabla_\sigma  A_{\lambda\tau} &= -
2K_{(\mu}^{\phantom{(\mu}\rho} D_{\nu)} A_{\rho\beta} - D_\mu
K_{\nu\rho}A^{\rho}_{\phantom\rho\beta} + K_{\mu\nu}a^\rho A_{\rho\beta}
\,,\\
-n_\beta &&
\projector{\rho}{\mu}\projector{\sigma}{\nu}\projector{\lambda}{\alpha}
n^{\tau}\nabla_\rho \nabla_\sigma  A_{\lambda\tau} &= -
2K_{(\mu}^{\phantom{(\mu}\rho} D_{\nu)} A_{\alpha\rho} - D_\mu
K_{\nu\rho}A_\alpha^{\phantom\alpha\rho} + K_{\mu\nu}
A_{\alpha\rho}a^\rho \,,\\
+n_\mu n_\alpha &&
n^{\rho}\projector{\sigma}{\nu}n^{\lambda}\projector{\tau}{\beta}
\nabla_\rho \nabla_\sigma  A_{\lambda\tau} &= - K_\nu^{\phantom\nu\rho}
\frac{1}{N}\cL_{Nn}A_{\rho\beta} - \frac{1}{N}\cL_{Nn}
K_{\nu\rho}A^{\rho}_{\phantom\rho\beta} - a^\rho D_\nu A_{\rho\beta}
\nn\\
&&&+ 3K_\nu^{\phantom\nu\rho}K_\rho^{\phantom\rho\sigma}A_{\sigma\beta}
+ K_\nu^{\phantom\nu\rho}K_\beta^{\phantom\beta\sigma}A_{\rho\sigma} +
a_\nu a^\rho A_{\rho\beta} \,,\\
+n_\mu n_\beta &&
n^{\rho}\projector{\sigma}{\nu}\projector{\lambda}{\alpha}n^{\tau}
\nabla_\rho \nabla_\sigma  A_{\lambda\tau} &= - K_\nu^{\phantom\nu\rho}
\frac{1}{N}\cL_{Nn}A_{\alpha\rho} - \frac{1}{N}\cL_{Nn}
K_{\nu\rho}A_\alpha^{\phantom\alpha\rho} - a^\rho D_\nu A_{\alpha\rho}
\nn\\
&&&+ 3K_\nu^{\phantom\nu\rho}K_\rho^{\phantom\rho\sigma}A_{\alpha\sigma}
+ K_\nu^{\phantom\nu\rho}K_\alpha^{\phantom\alpha\sigma}A_{\rho\sigma} +
a_\nu a^\rho A_{\alpha\rho} \,,\\
+n_\nu n_\alpha &&
\projector{\rho}{\mu}n^{\sigma}n^{\lambda}\projector{\tau}{\beta}
\nabla_\rho \nabla_\sigma  A_{\lambda\tau} &= - K_\mu^{\phantom\mu\rho}
\frac{1}{N}\cL_{Nn}A_{\rho\beta} - a^\rho D_\mu A_{\rho\beta} - D_\mu
a_\rho A^{\rho}_{\phantom\rho\beta} \nn\\
&&&+ 2K_\mu^{\phantom\mu\rho}K_\rho^{\phantom\rho\sigma}A_{\sigma\beta}
+ K_\mu^{\phantom\mu\rho}K_\beta^{\phantom\beta\sigma}A_{\rho\sigma}
\,,\\
+n_\nu n_\beta &&
\projector{\rho}{\mu}n^{\sigma}\projector{\lambda}{\alpha}n^{\tau}
\nabla_\rho \nabla_\sigma  A_{\lambda\tau} &=  - K_\mu^{\phantom\mu\rho}
\frac{1}{N}\cL_{Nn}A_{\alpha\rho} - a^\rho D_\mu A_{\alpha\rho} - D_\mu
a_\rho A_\alpha^{\phantom\alpha\rho} \nn\\
&&&+ 2K_\mu^{\phantom\mu\rho}K_\rho^{\phantom\rho\sigma}A_{\alpha\sigma}
+ K_\mu^{\phantom\mu\rho}K_\alpha^{\phantom\alpha\sigma}A_{\rho\sigma}
\,,\\
+n_\alpha n_\beta &&
\projector{\rho}{\mu}\projector{\sigma}{\nu}n^{\lambda}n^{\tau}
\nabla_\rho \nabla_\sigma  A_{\lambda\tau} &= 2K_\mu^{\phantom\mu\rho}
K_\nu^{\phantom\nu\sigma} A_{\rho\sigma} \,,\\
-n_\mu n_\nu n_\alpha &&
n^{\rho}n^{\sigma}n^{\lambda}\projector{\tau}{\beta}\nabla_\rho
\nabla_\sigma  A_{\lambda\tau} &= -
2a^{\rho}\frac{1}{N}\cL_{Nn}A_{\rho\beta} - \frac{1}{N}\cL_{Nn} a_\rho
A^{\rho}_{\phantom\rho\beta} \nn\\
&&&+ 4a^{\rho}K_\rho^{\phantom\rho\sigma}A_{\sigma\beta} +
2a^{\rho}K_\beta^{\phantom\beta\sigma}A_{\rho\sigma} \,,\\
-n_\mu n_\nu n_\beta &&
n^{\rho}n^{\sigma}\projector{\lambda}{\alpha}n^{\tau}\nabla_\rho
\nabla_\sigma  A_{\lambda\tau} &=  -
2a^{\rho}\frac{1}{N}\cL_{Nn}A_{\alpha\rho} - \frac{1}{N}\cL_{Nn} a_\rho
A_\alpha^{\phantom\alpha\rho} \nn\\
&&&+ 4a^{\rho}K_\rho^{\phantom\rho\sigma}A_{\alpha\sigma} +
2a^{\sigma}K_\alpha^{\phantom\alpha\rho}A_{\rho\sigma} \,,\\
-n_\mu n_\alpha n_\beta &&
n^{\rho}\projector{\sigma}{\nu}n^{\lambda}n^{\tau}\nabla_\rho
\nabla_\sigma  A_{\lambda\tau} &= 2a^{\rho} K_\nu^{\phantom\nu\sigma}
A_{\rho\sigma} \,,\\
-n_\nu n_\alpha n_\beta &&
\projector{\rho}{\mu}n^{\sigma}n^{\lambda}n^{\tau}\nabla_\rho
\nabla_\sigma  A_{\lambda\tau} &= 2K_\mu^{\phantom\mu\rho} a^{\sigma}
A_{\rho\sigma} \,,\\
+n_\mu n_\nu n_\alpha n_\beta &&
n^{\rho}n^{\sigma}n^{\lambda}n^{\tau}\nabla_\rho \nabla_\sigma
A_{\lambda\tau} &= 2a^{\rho} a^{\sigma} A_{\rho\sigma} \,.
\end{align*}
Finally, we have obtained every component of the decomposition of
$\nabla_\mu \nabla_\nu A_{\alpha\beta}$.

One thing remains to be uncovered in the spacetime decomposition.
The Lie derivative \eqref{LieNnB} contains a time derivative of
$B_{\alpha_1 \cdots\alpha_k}$ that needs to be uncovered. For this
purpose we assume a coordinate system on $\Sigma_t$ and use the
connection coefficients explicitly. We denote the connection
coefficients of $\nabla$ and $D$ with
\begin{equation}
\GammaM^\mu_{\nu\rho}=\frac{1}{2}\gM^{\mu\sigma}\left( \p_\nu
\gM_{\sigma\rho}+\p_\rho \gM_{\nu\sigma}-\p_\sigma \gM_{\nu\rho} \right)
\end{equation}
and
\begin{equation}
\Gamma^i_{jk}=\frac{1}{2}g^{im}\left( \p_j g_{mk}+\p_k g_{jm}-\p_m
g_{jk} \right) \,,
\end{equation}
respectively. Using \eqref{ADMmetric} and \eqref{ADMinvmetric} we obtain
\begin{align}
\GammaM^0_{00} &= \frac{1}{N}\left( \dot{N}+N^i \p_i N + K_{ij}N^i N^j
\right) \,,\\
\GammaM^0_{0i} &= \frac{1}{N}\left( \p_i N + K_{ij}N^j \right) \,,\\
\GammaM^0_{ij} &= \frac{1}{N}K_{ij} \,,\\
\GammaM^i_{00} &= g^{ij} \left( N\p_j N + \dot{N}_j - D_j N_k N^k
\right) - \frac{N^i}{N}\left( \dot{N} + N^j \p_j N + K_{jk}N^j N^k
\right) \,,\\
\GammaM^i_{0j} &= g^{ik} \left( NK_{kj} + D_j N_k \right) -
\frac{N^i}{N}\left( \p_j N + N^k K_{kj} \right) \,,\\
\GammaM^i_{jk} &= \Gamma^i_{jk} - \frac{N^i K_{jk}}{N} \,.
\end{align}
Then we obtain the covariant derivative in \eqref{LieNnB} by using the
connection coefficients. For our purposes it is sufficient to consider
only the spatial components of \eqref{LieNnB} because the action
\eqref{S3.gf.ADM.2} involves only them \eqref{LieNnzeta}. We obtain
\begin{equation}
\begin{split}
\nabla_n B_{i_1 \cdots i_k} &= n^\mu \left( \p_\mu B_{i_1 \cdots i_k} -
\GammaM^\nu_{\mu i_1}B_{\nu \cdots i_k} - \cdots - \GammaM^\nu_{\mu
i_k}B_{i_1 \cdots \nu} \right)\\
&= \frac{1}{N}\left[ \dot{B}_{i_1 \cdots i_k} - N^j D_j B_{i_1 \cdots
i_k} - \left( NK_{i_1}^{\phantom{i_1}j}+D_{i_1}N^j \right) B_{j \cdots
i_k} \right.\\
&\qquad\qquad - \cdots - \left( NK_{i_k}^{\phantom{i_k}j}+D_{i_k}N^j
\right) B_{i_1 \cdots j} \Bigr]
\end{split}
\end{equation}
Thus we can write the spatial components of \eqref{LieNnB} as
\begin{equation}
\begin{split}
\frac{1}{N}\cL_{Nn}B_{i_1 \cdots i_k} &= \frac{1}{N}\left(  \dot{B}_{i_1
\cdots i_k} - N^j D_j B_{i_1 \cdots i_k} - D_{i_1}N^jB_{j \cdots i_k} -
\cdots - D_{i_k}N^j B_{i_1 \cdots j} \right)\\
&= \frac{1}{N}\left(  \dot{B}_{i_1 \cdots i_k} - \cL_{\bm{N}} B_{i_1
\cdots i_k} \right) \,,
\end{split}
\end{equation}
where $\cL_{\bm{N}}$ denotes the Lie derivative along the shift vector
$N^i$ in $\Sigma_t$.

\subsection{Construction of the inverse to
\texorpdfstring{$F^{ijkl}$}{F(ijkl)}}\label{appendix4}
Here we shall obtain the inverse to the tensor $F^{ijkl}$ defined in
\eqref{F^ijkl}. First we note that $F^{ijkl}$ is symmetric in its first
two indices and also in its last two indices:
\begin{equation}\label{Fsymmetries}
F^{ijkl}=F^{jikl} \,,\qquad F^{ijkl}=F^{ijlk} \,.
\end{equation}
It also has the properties
\begin{align*}
g_{ij}F^{ijkl} &= g^{kl} \,,\\
g_{kl}F^{ijkl} &= g^{ij} - 8\kappa^2\alpha U_0 \left(
g^{ik}g^{jl}-\frac{1}{3}g^{ij}g^{kl} \right) \zeta_{km}\zeta_{ln}g^{mn}
\,,
\end{align*}
where the first result is directly related to the fact that both sides
of \eqref{solve:barK_ij} are traceless.

In order to prove that the inverse $F^{-1}_{ijkl}$ to $F^{ijkl}$
actually exists, and consequently that no more primary constraints are
required, we shall construct $F^{-1}_{ijkl}$.
First note that $F^{-1}_{ijkl}$ must have similar symmetries
\eqref{Fsymmetries} as $F^{ijkl}$:
\begin{equation}\label{invFsymmetries}
F^{-1}_{ijkl}=F^{-1}_{jikl}\,,\qquad F^{-1}_{ijkl}=F^{-1}_{ijlk}\,.
\end{equation}
We can define $F^{-1}_{ijkl}$ as a power series
\begin{equation}\label{invF}
\begin{split}
F^{-1}_{ijkl} = \frac{1}{2}\left( g_{ik}g_{jl}+g_{il}g_{jk} \right)
+ &\sum_{J=1}^\infty \left( 8\kappa^2\alpha U_0 \right)^J
C^{(J)m_1n_1\cdots m_Jn_J}_{ijkl} \\
&\times \zeta_{m_1o_1}\zeta_{n_1p_1}g^{o_1p_1}\cdots
\zeta_{m_Jo_J}\zeta_{n_Jp_J}g^{o_Jp_J}\,,
\end{split}
\end{equation}
where the coefficients $C^{(J)m_1n_1\cdots m_Jn_J}_{ijkl}$ depend only
on the spatial metric and possess the symmetries \eqref{invFsymmetries}
of $F^{-1}_{ijkl}$ with respect to the indices $ij$ and $kl$.
These coefficients are defined by the identity
$F^{-1}_{ijkl}F^{klmn}=\delta_{i}^{(m}\delta_{j}^{n)}$ as follows.
The first coefficient is
\begin{equation*}
C^{(1)mn}_{ijkl} = \delta_{(i}^m g_{j)(k}\delta_{l)}^n
- \frac{1}{3}g_{ij}\delta_{(k}^m\delta_{l)}^n
\end{equation*}
and the rest of the coefficients are defined recursively by the formula
\begin{equation}\label{coefficientC}
C^{(J)m_1n_1\cdots m_Jn_J}_{ijkl} =
C^{(J-1)m_1n_1\cdots m_{J-1}n_{J-1}}_{ijop}
\left( g^{m_J(o}\delta^{p)}_{(k}\delta_{l)}^{n_J}
- \frac{1}{3}g^{op}\delta_{(k}^{m_J}\delta_{l)}^{n_J} \right)
\end{equation}
for every order $J>1$.
The coefficient $C^{(J)m_1n_1\cdots m_Jn_J}_{ijkl}$ of order $J$ in
\eqref{coefficientC} vanishes if the coefficient $C^{(J-1)m_1n_1\cdots
m_{J-1}n_{J-1}}_{ijkl}$ of order $J-1$ is proportional to $g_{kl}$.
In that case the series would be truncated.
However, it seems that such a coefficient does not appear and hence the
recursion \eqref{coefficientC} will proceed \emph{ad infinitum} for
coefficients with increasingly complicated index configurations.
For example, the second-order coefficient is
\begin{multline*}
C^{(2)m_1n_1m_2n_2}_{ijkl} = \frac{1}{2}\delta_{(i}^{m_1}
\delta_{j)}^{m_2}\delta_{(k}^{n_1}\delta_{l)}^{n_2}
+ \frac{1}{2}\delta_{(i}^{m_1}g_{j)(k}\delta_{l)}^{n_2}g^{n_1m_2}
- \frac{1}{3}\delta_{(i}^{m_1}\delta_{j)}^{n_1}
\delta_{(k}^{m_2}\delta_{l)}^{n_2}\\
- \frac{1}{3}g_{ij}\left(\frac{1}{2}g^{m_1m_2}
\delta_{(k}^{n_1}\delta_{l)}^{n_2}
+ \frac{1}{2}g^{n_1m_2}\delta_{(k}^{m_1}\delta_{l)}^{n_2}
-\frac{1}{3}g^{m_1n_1}\delta_{(k}^{m_2}\delta_{l)}^{n_2}
 \right)\,.
\end{multline*}
Note that for every order $J$ we obtain
\begin{equation*}
g^{ij}C^{(J)m_1n_1\cdots m_Jn_J}_{ijkl} = 0\,,
\end{equation*}
which implies
\begin{equation}\label{gF^-1}
g^{ij}F^{-1}_{ijkl} = g_{kl}\,.
\end{equation}

For $F^{-1}_{ijkl}$ to be defined as the infinite series \eqref{invF}
every one of its components has to converge. Consider the series
$\sum_{J=1}^\infty a_J$ with terms
\begin{equation*}
a_J=\left( 8\kappa^2\alpha U_0 \right)^J
A^{ij}C^{(J)m_1n_1\cdots m_Jn_J}_{ijkl}
\zeta_{m_1o_1}\zeta_{n_1p_1}g^{o_1p_1}\cdots
\zeta_{m_Jo_J}\zeta_{n_Jp_J}g^{o_Jp_J}B^{kl}\,,
\end{equation*}
where $A^{ij}$ and $B^{ij}$ are arbitrary symmetric tensors.
Convergence of this series would clearly imply the convergence of
every component of $F^{-1}_{ijkl}$, because
\begin{equation*}
A^{ij}F^{-1}_{ijkl}B^{kl}=A_{ij}B^{ij}+\sum_{J=1}^\infty a_J
\end{equation*}
for any $A^{ij}$ and $B^{ij}$. Since the terms $a_J$ may have varying
signs let us consider absolute convergence, i.e., convergence of the
series $\sum_{J=1}^\infty |a_J|$, which always implies the convergence
of the original series. The ratio of successive terms in the series is
\begin{equation}\label{a_J.ratio}
\frac{|a_{J+1}|}{|a_J|}=|8\kappa^2\alpha U_0|
\frac{\left|\cA^{(J)}_{kl}\left(g^{m(k}B^{l)n}-\frac{1}{3}g^{kl}B^{mn}
\right)\zeta_{mo}\zeta_{np}g^{op}\right|}{\left|\cA^{(J)}_{kl}B^{kl}
\right|}
\end{equation}
where we denote
\begin{equation*}
\cA^{(J)}_{kl}=A^{ij}C^{(J)m_1n_1\cdots m_Jn_J}_{ijkl}
\zeta_{m_1o_1}\zeta_{n_1p_1}g^{o_1p_1}\cdots
\zeta_{m_Jo_J}\zeta_{n_Jp_J}g^{o_Jp_J}\,.
\end{equation*}
For any given range of values for the components of the tensors
$\zeta_{ij}$ and $g_{ij}$ (and its inverse $g^{ij}$) it should be
possible to choose a sufficiently small coupling $|\alpha|$ so that the
ratio \eqref{a_J.ratio} stays smaller than 1 for sufficiently high $J$,
and eventually when $J\rightarrow\infty$. Of course here both
$\zeta_{ij}$ and $g_{ij}$ are dynamical field variables and cannot be
restricted to any given range of values \emph{a priori}. However, for
any given geometry of spacetime there should exist sufficiently small
couplings $|\alpha|$ that ensure absolute convergence of the series
\eqref{invF}. This shows that the series may converge, but not always.
Admittedly, convergence is bound to limit the coupling constant,
depending on the geometry of spacetime and Hamiltonian structure. This
can be a hint that the CRG action is not a consistent one for arbitrary
coupling constants. Physically, the reason for the appearance of such
an infinite series in the Hamiltonian structure of CRG is not quite
clear. Infinite series appear, e.g. in the Hamiltonian structure of
nonlocal theories, albeit they are quite different since they
typically involve an infinite number of variables, whereas here the
series contains only the two fields $g_{ij}$ and $\zeta_{ij}$ and no
momenta.

\subsection{Consistency of secondary constraints in time}
\label{appendix5}
Here we shall ensure that every secondary constraint is preserved in
time. First we consider the constraints $\cH_i$. We introduce a
global smeared version of the constraint $\cH_i$ as
\begin{equation}\label{Phi_S}
\Phi_S(\eta^i) = \int d^3\bx \eta^i \cH_i \,,
\end{equation}
where $\eta^i$ ($i=1,2,3$) are arbitrary functions on $\Sigma_t$ which
vanish rapidly enough at infinity.
The Poisson brackets of the constraint \eqref{Phi_S} with the canonical
variables are
\begin{align*}
\pb{ \Phi_S(\eta^k), g_{ij} } &= - \eta^k \p_k g_{ij} - \p_i \eta^k
g_{kj} - \p_j \eta^k g_{ki} = -\cL_{\bm\eta}g_{ij} \,,\\
\pb{ \Phi_S(\eta^k), p^{ij} } &= - \p_k \eta^k p^{ij} - \eta^k \p_k
p^{ij} + \p_k \eta^i p^{kj} + \p_k \eta^j p^{ki}
 = -\cL_{\bm\eta}p^{ij} \,,\\
\pb{ \Phi_S(\eta^k), \zeta_{ij} } &= - \eta^k \p_k \zeta_{ij} - \p_i
\eta^k \zeta_{kj} - \p_j \eta^k \zeta_{ki}
= -\cL_{\bm\eta}\zeta _{ij} \,,\\
\pb{ \Phi_S(\eta^k), p_\zeta^{ij} } &= - \p_k \eta^k p_\zeta^{ij} -
\eta^k \p_k p_\zeta^{ij} + \p_k \eta^i p_\zeta^{kj} + \p_k \eta^j
p_\zeta^{ki} = -\cL_{\bm\eta}p_\zeta^{ij} \,.
\end{align*}
The constraints \eqref{Phi_S} also satisfy the diffeomorphism algebra
\begin{equation*}
\pb{ \Phi_S(\eta^i), \Phi_S(\theta^i) } =
\Phi_S(\eta^j\p_j\theta^i-\theta^j\p_j\eta^i) \,.
\end{equation*}
Thus we identify \eqref{Phi_S} as the momentum constraint that generates
diffeomorphisms in the spatial hypersurface $\Sigma_t$ for the dynamical
variables $g_{ij}$, $p^{ij}$, $\zeta_{ij}$ and $p_\zeta^{ij}$.
In fact we can extend the momentum constraint  \eqref{Phi_S} to a full
generator of spatial diffeomorphisms with the help of the primary
constraints \eqref{pconstraints}. We redefine
\begin{equation}\label{Phi_S.full}
\begin{split}
\Phi_S(\eta^i) &= \int d^3\bx \left( p_i\cL_{\bm\eta}N^i
+p^{ij}\cL_{\bm\eta}g_{ij}+p_\zeta^{ij}\cL_{\bm\eta}\zeta _{ij}
+p^\lambda_{ij}\cL_{\bm\eta}\lambda^{ij} \right)\\
&= \int d^3\bx \eta^i \Phi_i \,,
\end{split}
\end{equation}
where
\begin{equation}\label{Phi_i}
\Phi_i = \cH_i + \cL_{\bm N}p_i + 2\p_j p^\lambda_{ik}\lambda^{jk}
+ p^\lambda_{jk}\p_i\lambda^{jk} + 2p^\lambda_{ik}\p_j\lambda^{jk}\,.
\end{equation}
Note that the generator for $N$ and $p_N$ vanishes,
\[
\int d^3\bx
p_N\cL_{\bm\eta}N=\int d^3\bx\eta^i p_N\p_iN=0\,,
\]
since these variables are spatial constants.
The extension \eqref{Phi_S.full} of the momentum constraint simply
amounts to rewriting the Lagrange multipliers of the primary
constraints \eqref{pconstraints} as
\begin{equation*}
u^i=v^i+\cL_{\bm N}N^i \,,\qquad
u_\lambda^{ij}=v_\lambda^{ij}+\cL_{\bm N}\lambda^{ij}\,,\qquad
\int d^3\bx u_N=v_N\,,
\end{equation*}
where $v^i$, $v_\lambda^{ij}$, and $v_N$ are arbitrary.
The constraint \eqref{Phi_S.full} evidently generates time-dependent
spatial diffeomorphisms for all the variables.
Therefore we obtain
\begin{equation*}
\pb{\Phi_S(\eta^i),\cH_0}=
-\eta^i\partial_i\cH_0-\partial_i\eta^i\cH_0 \,.
\end{equation*}
In other words, $\cH_0$ is a scalar density on the spatial hypersurface
$\Sigma_t$.
We again emphasize that the lapse $N$ depends only on time.
As a result there is no local Hamiltonian constraint $\cH_0$ but only
the global one $\Phi_0$ defined in \eqref{Phi_0}.
As a result we clearly have
\begin{align}
\pb{\Phi_0,\Phi_0}&=0 \,,\label{pb:Phi_0,Phi_0}\\
\pb{\Phi_S(\eta^i),\Phi_0}&=0 \,.\label{pb:Phi_S,Phi_0}
\end{align}
It is now evident that the momentum constraint \eqref{Phi_i} is
preserved in time, since \eqref{pb:Phi_0,Phi_0} and
\eqref{pb:Phi_S,Phi_0} and all the rest of the constraints in the
Hamiltonian transform as scalar or tensor densities under spatial
diffeomorphism.
Indeed we can write the Hamiltonian as a sum of the constraints
\begin{equation}\label{H.2nd}
H=N\Phi_0+\Phi_S(N^i) + v_N p_N +\int d^3\bx \left( v^i p_i +
v_\lambda^{ij}p^\lambda_{ij}+\bar{v}_{ij}\bar{\Pi}^{ij} \right).
\end{equation}
We could also include the 11 secondary constraints $\Psi_{ij}$ and
$\bar{\Pi}_{II}^{ij}$ into the Hamiltonian with arbitrary Lagrange
multipliers. However, those 11 Lagrange multipliers would already be
fixed by the 11 consistency conditions for the primary constraints
$p^\lambda_{ij}$ and $\bar{\Pi}^{ij}$. Thus the inclusion of those
constraints is of no benefit to us. Hence we leave them out from the
Hamiltonian.

The global Hamiltonian constraint $\Phi_0$ is preserved in time
due to the constraints $p^\lambda_{ij}$, $\Psi_{ij}$, $\bar{\Pi}^{ij}$,
$\bar{\Pi}_{II}^{ij}$:
\begin{equation*}
\pb{\Phi_0,H}\approx -\int d^3\bx\left( v_\lambda^{ij}\Psi_{ij}
+\bar{v}_{ij}\bar{\Pi}_{II}^{ij} \right) \approx 0\,.
\end{equation*}

Next we have to ensure that the secondary constraint $\Psi_{ij}$ is
preserved in time. The Poisson brackets $\pb{\bar{\Psi}_{ij},H}$ and
$\pb{\Psi,H}$ must both be either zero or a constraint.
The consistency conditions can be written as
\begin{equation}\label{eq.v_xi}
N\pb{\bar{\Psi}_{ij}(\bx),\Phi_0} +
\int d^3\by\left( v_\lambda^{kl}(\by)
\pb{\bar{\Psi}_{ij}(\bx),p^\lambda_{kl}(\by)}
+ \bar{v}_{kl}(\by)\pb{\bar{\Psi}_{ij}(\bx),\bar{\Pi}^{kl}(\by)}
\right)=0
\end{equation}
and
\begin{equation}\label{pb:Psi,H}
N\pb{\Psi(\bx),\Phi_0}+3\alpha\sqrt{g}
\frac{\bar{v}_{ij}\bar{\zeta}^{ij}}{\zeta}(\bx)=0
\end{equation}
for the traceless component $\bar{\Psi}_{ij}$ and the trace
component $\Psi$, respectively.
In \eqref{pb:Psi,H}, we used \eqref{pb:Psi,p^xi_kl} and
\begin{equation}\label{pb:Psi,barPi}
\pb{\Psi(\bx),\bar{\Pi}^{ij}(\by)}=3\alpha\sqrt{g}
\frac{\bar{\zeta}^{ij}}{\zeta}(\bx) \delta(\bx-\by) \,.
\end{equation}
The consistency condition \eqref{pb:Psi,H} can be solved for one of
the five independent components in the Lagrange multiplier
$\bar{v}_{ij}$. Let us denote the specific solution to \eqref{pb:Psi,H}
by $\bar{v}'_{ij}$, which still contains four independent arbitrary
components. The solved component of $\bar{v}'_{ij}$ is proportional to
$N$. The homogeneous part of \eqref{pb:Psi,H} has only the nontrivial
solution. Consistency of $\Psi$ is assured by substituting the solution
$\bar{v}_{ij}=\bar{v}'_{ij}$ into the Hamiltonian.
Then the consistency of the constraint $\bar{\Psi}_{ij}$ is assured by
solving the inhomogeneous linear equation \eqref{eq.v_xi} for the
Lagrange multiplier $v_\lambda^{kl}$. None of the three Poisson brackets
in
\eqref{eq.v_xi} vanishes, not even weakly.
Explicitly the second Poisson bracket can be written as
\begin{equation}\label{pb:Psi_ij,p^xi_kl}
\begin{split}
\pb{\bar{\Psi}_{ij}(\bx),p^{\lambda}_{kl}(\by)}=-\frac{1}{\sqrt{g}}
\frac{\kappa^2}{4U_0^2} &\left(
A_{ijkl}-\frac{1}{3}g_{ij}g^{mn}A_{mnkl} \right.\\
&+\left. F^{-1}_{ijkl}-\frac{1}{3} g_{kl}F^{-1}_{ijmn}g^{mn}\right)
\left(\frac{p_\zeta}{\zeta}\right)^2(\bx)
\delta(\bx-\by)\,,
\end{split}
\end{equation}
where we denote
\begin{equation*}
\begin{split}
A_{ijkl}&=F^{-1}_{klij} - F^{-1}_{mnij}g^{mo}g^{np}\left(
F^{-1}_{opkl}-\frac{1}{3}g_{kl}F^{-1}_{opqr}g^{qr} \right)\\
&\quad+8\kappa^2\alpha U_0 \zeta^{mn}F^{-1}_{moij}\zeta^p_{\phantom{p}n}
g^{oq}\left( F^{-1}_{pqkl}-\frac{1}{3}g_{kl}F^{-1}_{pqrs}g^{rs}
\right) \,.
\end{split}
\end{equation*}
Thus we obtain
\begin{equation*}
g^{kl}(\by)\pb{\bar{\Psi}_{ij}(\bx),p^\lambda_{kl}(\by)}=0\,,
\end{equation*}
where we also used the property \eqref{gF^-1}.
Clearly the homogeneous part of the equation \eqref{eq.v_xi}
\begin{equation}\label{eq.v_xi.homog}
\int d^3\by v_\lambda^{kl}(\by)
\pb{\bar{\Psi}_{ij}(\bx),p^\lambda_{kl}(\by)}=0
\end{equation}
has the solution
\begin{equation}\label{v_xi.homog}
v_\lambda^{kl}=\frac{1}{3}v_\lambda g ^{kl}\,,
\end{equation}
where the trace component $v_\lambda$ of the Lagrange multiplier
$v_\lambda^{kl}$ is still an arbitrary field. This is the only
nontrivial
solution to the homogeneous equation \eqref{eq.v_xi.homog}.
The specific solution to the inhomogeneous equation \eqref{eq.v_xi} has
the form
\begin{equation}\label{Nbarw_xi}
v_\lambda^{kl}=N\bar{w}_\lambda^{kl}\bigl[g_{ij},p^{ij},\zeta_{ij},
p_\zeta,
\lambda^{ij},\bar{v}'_{ij}/N\bigr] \,,
\end{equation}
where $\bar{w}_\lambda^{kl}$ is a traceless functional of the listed
variables and of the Lagrange multiplier $\bar{v}'_{ij}$ divided by
$N$. Formally it can be written as
\begin{equation}\label{barw_xi}
\begin{split}
\bar{w}_\lambda^{ij}(\bx)&=-\int d^3\by
B^{ijkl}(\bx,\by)\pb{\bar{\Psi}_{kl}(\by),\Phi_0}\\
&\quad-\iint d^3\by d^3\bz \frac{\bar{v}'_{kl}(\bz)}{N}
B^{ijmn}(\bx,\by)\pb{\bar{\Psi}_{mn}(\by),\bar{\Pi}^{kl}(\bz)}\,,
\end{split}
\end{equation}
where $B^{ijkl}(\bx,\by)$ is the inverse to \eqref{pb:Psi_ij,p^xi_kl},
i.e., it satisfies
\begin{equation*}
\int d^3\by
B^{ijkl}(\bx,\by)\pb{\bar{\Psi}_{kl}(\by),p^{\lambda}_{mn}(\bz)}
=\delta^{(i}_m\delta^{j)}_n \delta(\bx-\bz)\,.
\end{equation*}
The general solution to \eqref{eq.v_xi} is the sum of the specific
solution \eqref{Nbarw_xi} and the solution \eqref{v_xi.homog} to
the homogeneous equation \eqref{eq.v_xi.homog}:
\begin{equation*}
v_\lambda^{kl}=N\bar{w}_\lambda^{kl}\bigl[g_{ij},p^{ij},\zeta_{ij},
p_\zeta,
\lambda^{ij},\bar{v}'_{ij}/N\bigr]+ \frac{1}{3}v_\lambda g^{kl}\,.
\end{equation*}
Inserting this solution into the Hamiltonian \eqref{H.2nd} ensures
the consistency of $\bar{\Psi}_{ij}$.
The Hamiltonian is now written as
\begin{equation}\label{H.3rd}
H=N\Phi_0+N\int
d^3\bx\bar{w}_\lambda^{ij}\bar{p}^\lambda_{ij}+\Phi_S(N^i)
+v_N p_N +\int d^3\bx \left( v^i p_i +\frac{1}{3}v_\lambda p^\lambda
+\bar{v}'_{ij}\bar{\Pi}^{ij} \right).
\end{equation}
Lagrange multipliers that are arbitrary or at least contain some
arbitrary components are denoted by $v$ and the multipliers denoted by
$w$ have been solved entirely in order to ensure the consistency of
constraints, e.g., in \eqref{H.3rd} $\bar{w}_\lambda^{ij}$ is given by
the
specific solution \eqref{barw_xi} to \eqref{eq.v_xi} and $\bar{v}'_{ij}$
is the solution to \eqref{pb:Psi,H} with four arbitrary components left.

As a last step in Dirac's algorithm we have to ensure the consistency of
the secondary constraints $\bar{\Pi}_{II}^{ij}$ defined in
\eqref{barPi_II}. We require that
\begin{multline}\label{pb:barPi_II,H}
\pb{\bar{\Pi}_{II}^{ij}(\bx),H}\approx
N\pb{\bar{\Pi}_{II}^{ij}(\bx),\Phi_0}
+N\int d^3\by \bar{w}_\lambda^{kl}(\by)
\pb{\bar{\Pi}_{II}^{ij}(\bx),\bar{p}^\lambda_{kl}(\by)} \\
+\int d^3\by \frac{1}{3}v_\lambda(\by)
\pb{\bar{\Pi}_{II}^{ij}(\bx),p^\lambda(\by)}
+\int d^3\by \bar{v}'_{kl}(\by)
\pb{\bar{\Pi}_{II}^{ij}(\bx),\bar{\Pi}^{kl}(\by)}
\end{multline}
must be either zero or a constraint. In this equation
$\bar{w}_\lambda^{kl}$
is given by \eqref{barw_xi}.
For the calculation of the Poisson brackets of $\bar{\Pi}_{II}^{ij}$
with $p^\lambda$ and $\bar{p}^\lambda_{kl}$ we can use the
definition of $\bar{\Pi}_{II}^{ij}(\bx)$ as the Poisson
bracket $\pb{\bar{\Pi}^{ij}(\bx),\Phi_0}$, and the Jacobi identity for
the Poisson bracket. First we obtain
\begin{equation*}
\begin{split}
\pb{\bar{\Pi}_{II}^{ij}(\bx),p^\lambda_{kl}(\by)} &=
-\pb{\pb{\Phi_0,p^\lambda_{kl}(\by)},\bar{\Pi}^{ij}(\bx)}
-\pb{\pb{p^\lambda_{kl}(\by),\bar{\Pi}^{ij}(\bx)},\Phi_0}\\
&=\pb{\Psi_{kl}(\by),\bar{\Pi}^{ij}(\bx)}\\
&=\pb{\bar{\Psi}_{kl}(\by),\bar{\Pi}^{ij}(\bx)}
+\frac{1}{3}g_{kl}(\by)\pb{\Psi(\by),\bar{\Pi}^{ij}(\bx)} \,,
\end{split}
\end{equation*}
where we also used the independence of $\bar{\Pi}^{ij}$ on
$\lambda^{ij}$.
For the trace component $p^\lambda$ we obtain
\begin{equation*}
\begin{split}
\pb{\bar{\Pi}_{II}^{ij}(\bx),p^\lambda(\by)}
&=-\pb{\pb{\Phi_0,p^\lambda(\by)},\bar{\Pi}^{ij}(\bx)}
-\pb{\pb{p^\lambda(\by),\bar{\Pi}^{ij}(\bx)},\Phi_0}\\
&=-\pb{\pb{\Phi_0,g^{kl}(\by)},\bar{\Pi}^{ij}(\bx)}p^\lambda_{kl}(\by)
+\pb{\Psi(\by),\bar{\Pi}^{ij}(\bx)} \\
&\approx \pb{\Psi(\by),\bar{\Pi}^{ij}(\bx)} \,,
\end{split}
\end{equation*}
where the last expression is given in \eqref{pb:Psi,barPi}.
For the traceless component $\bar{p}^\lambda_{ij}$ we obtain
\begin{equation*}
\begin{split}
\pb{\bar{\Pi}_{II}^{ij}(\bx),\bar{p}^\lambda_{kl}(\by)}
&= \pb{\bar{\Pi}_{II}^{ij}(\bx),p^\lambda_{kl}(\by)}
-\frac{1}{3}\pb{\bar{\Pi}_{II}^{ij}(\bx),g_{kl}(\by)}p^\lambda(\by)\\
&\quad-\frac{1}{3}g_{kl}(\by)
\pb{\bar{\Pi}_{II}^{ij}(\bx),p^\lambda(\by)}\\
&\approx \pb{\bar{\Psi}_{kl}(\by),\bar{\Pi}^{ij}(\bx)} \,,
\end{split}
\end{equation*}
which is a complicated nonvanishing expression.
The first and the last Poisson brackets in \eqref{pb:barPi_II,H} are
neither zero nor a combination of the constraints. Indeed they are
very complicated expressions. Thus, we see that the five consistency
conditions for $\bar{\Pi}_{II}^{ij}$ should be solved for the Lagrange
multiplier $v_\lambda$ and the four arbitrary Lagrange multipliers left
in $\bar{v}'_{kl}$. We obtain the following equations for them:
\begin{multline}\label{eq.restofLM}
N\pb{\bar{\Pi}_{II}^{ij}(\bx),\Phi_0}
+N\iint d^3\by d^3\bz\pb{\bar{\Pi}^{ij}(\bx),\bar{\Psi}_{kl}(\by)}
B^{klmn}(\by,\bz)\pb{\bar{\Psi}_{mn}(\bz),\Phi_0} \\
+\iiint d^3\by d^3\bz d^3\bz' \bar{v}'_{kl}(\bz')
\pb{\bar{\Pi}^{ij}(\bx),\bar{\Psi}_{mn}(\by)}
B^{mnop}(\by,\bz)\pb{\bar{\Psi}_{op}(\bz),\bar{\Pi}^{kl}(\bz')} \\
+\alpha\sqrt{g}\frac{\bar{\zeta}^{ij}}{\zeta}v_\lambda(\bx)
+\int d^3\by \bar{v}'_{kl}(\by)
\pb{\bar{\Pi}_{II}^{ij}(\bx),\bar{\Pi}^{kl}(\by)}=0\,.
\end{multline}
The general solution to \eqref{eq.restofLM} has the form
\begin{equation}\label{v_xi}
\begin{split}
v_\lambda &=
Nw\bigl[g_{ij},p^{ij},\zeta_{ij},p_\zeta,\lambda^{ij}\bigr]\,,\\
\bar{v}'_{kl} &=
N\bar{w}_{kl}\bigl[g_{ij},p^{ij},\zeta_{ij},p_\zeta,\lambda^{ij}\bigr]
+f\bar{h}_{kl}\bigl[g_{ij},p^{ij},\zeta_{ij},p_\zeta,\lambda^{ij}\bigr]\
,,
\end{split}
\end{equation}
where $w$ and $\bar{w}_{kl}$ comprise the specific solution to
\eqref{eq.restofLM} (divided by $N$), $\bar{h}_{kl}$ is a
possible nontrivial solution to the part of \eqref{eq.restofLM} that is
homogeneous in $\bar{v}'_{kl}$, and $f$ is an arbitrary functions of
time. Because of the very complicated form of Eq.~\eqref{eq.restofLM},
we have not been able to rigorously establish the
existence of neither the specific solution nor the homogeneous
solutions. Still we assume that a specific solution exists.
It is unclear whether a nontrivial solution $\bar{h}_{kl}$ to the
homogeneous part of \eqref{eq.restofLM} exists. Conceivably there could
even be several solutions to the homogeneous part.
The existence of a solution to the homogeneous equation would
imply the existence of an extra gauge symmetry associated with an
integrated (global) first-class constraint, a generator of the global
gauge transformation. We, however, suspect that such a global symmetry
does not exist in the theory, because no linear combination of the
constraints $\bar{\Pi}^{ij}$ appears to generate such symmetry.
Nevertheless if homogeneous solutions do exist they can be dealt with by
introducing some global gauge fixing conditions. In any case such
integrated constraints do not affect local dynamics. Hence in
Sec.~\ref{sec6}, we shall perform our analysis as if no nontrivial
solution to the part of \eqref{eq.restofLM} that is homogeneous in
$\bar{v}'_{kl}$ exists, i.e., we assume $\bar{h}_{kl}=0$.

\end{document}